\documentstyle[seceq,epsf,subfigure,wrapfig]{ptptex}

\newcommand{\qed}{\hbox{\rule[-2pt]{6pt}{6pt}}}

\makeatletter
\@ifundefined{lesssim}{}{}
\@ifundefined{gtrsim}{}{}
\def\vereq#1#2{\lower3pt\vbox{\baselineskip1.5pt \lineskip1.5pt
\ialign{$\m@th#1\hfill##\hfil$\crcr#2\crcr\sim\crcr}}}
\makeatother

\preprintnumber[3cm]{
WU-AP/160/03}

\title{
A Classification of Spherically Symmetric Kinematic Self-Similar Perfect-Fluid Solutions. II
}

\author{
Hideki~{\sc Maeda},$^{1,}$\footnote{E-mail: hideki@gravity.phys.waseda.ac.jp}
Tomohiro~{\sc Harada},$^{2,}$\footnote{E-mail: harada@gravity.phys.waseda.ac.jp}
Hideo~{\sc Iguchi}$^{3,}$\footnote{E-mail: iguchi@th.phys.titech.ac.jp}
\\ and \\
Naoya~{\sc Okuyama}$^{4,}$\footnote{E-mail: okuyama@gravity.phys.waseda.ac.jp}
}

\inst{
$^{1}$Advanced Research Institute for Science and Engineering,
Waseda University, Tokyo 169-8555, Japan
\\
$^{2,4}$Department of Physics, Waseda University, Tokyo 169-8555, Japan
\\
$^{3}$Department of Physics, Tokyo Institute of Technology, Tokyo 152-8550, Japan
}

\recdate{
}

\abst{
We give a classification of spherically symmetric kinematic self-similar solutions. This classification is complementary to that given in a previous work by the present authors [Prog.~Theor.~Phys.~{\bf 108},~819~(2002)]. Dust solutions of the second, zeroth and infinite kinds, perfect-fluid solutions and vacuum solutions of the first kind are treated. The kinematic self-similarity vector is either parallel or orthogonal to the fluid flow in the perfect-fluid and vacuum cases, while the `tilted' case, i.e., neither parallel nor orthogonal case, is also treated in the dust case. In the parallel case, there are no dust solutions of the second (except when the self-similarity index $\alpha$ is $3/2$), zeroth and infinite kinds, and in the orthogonal case, there are no dust solutions of the second and infinite kinds. Except in these cases, the governing equations can be integrated to give exact solutions. It is found that the dust solutions in the tilted case belong to a subclass of the Lema{\^ i}tre-Tolman-Bondi family of solutions for the marginally bound case. The flat Friedmann-Robertson-Walker (FRW) solution is the only dust solution of the second kind with $\alpha=3/2$ in the tilted and parallel cases and of the zeroth kind in the orthogonal case. The flat, open and closed FRW solutions with $p=-\mu/3$, where $p$ and $\mu$ are the pressure and energy density, respectively, are the only perfect-fluid first-kind self-similar solutions in the parallel case, while a new exact solution with $p=\mu$, which we call the ``singular stiff-fluid solution'', is the only such solution in the orthogonal case. The Minkowski solution is the only vacuum first-kind self-similar solution both in the parallel and orthogonal cases. Some important corrections and complements to the authors' previous work are also presented.}

\begin{document}

\maketitle

\section{Introduction}
Scale-invariance is one of the most fundamental properties of classical gravity. In the vacuum case without a cosmological constant, the gravitational constant $G$ is the only dimensional constant that appears in the field equations in Newtonian gravity, while there appears another physical constant, the speed of light, $c$, in general relativity. Neither a characteristic time nor a length scale can be constructed from them. This implies that there exists a class of scale-invariant solutions, which are often referred to as self-similar solutions. There also exist self-similar solutions in scalar-tensor theories of gravity. (See Ref.~\citen{mkm2002} for an investigation of the perfect-fluid case in Brans-Dicke theory.) In contrast, there may not exist self-similar solutions in quantum gravity, since, by definition, there appears a third dimensional constant, the Planck constant, $h$, with dimensions independent of that of $G$ and $c$. Matter fields that include some dimensional constants could also lead to violation of the scale-invariance. In general relativity, matter fields consistent with self-similarity include no constants with dimensions independent of those of $G$ and $c$. 

Self-similar solutions have been studied extensively, especially in spherically symmetric systems. This is not only because the self-similarity ansatz leads to the reduction of the governing equations from partially differential equations to ordinary differential equations, whose mathematical treatment is generally simpler, but also because there exist many applications. Although self-similar solutions are only special solutions of the field equations, they often play an important role as an attractor or an intermediate attractor. In particular, there is a {\it self-similarity hypothesis} which states that solutions in a variety of astrophysical and cosmological situations may naturally evolve into self-similar forms even if they are initially more complicated.~\cite{carr1999} In recent years, the validity of this hypothesis in spherically symmetric gravitational collapse has been reported both in Newtonian gravity~\cite{larson1969,penston1969,fc1993,hn1997,ti1999,mh2001,hms2003} and general relativity.~\cite{op1987,op1988,op1990,harada1998,hm2001,harada2001} 

Self-similarity in general relativity is defined by a self-similarity generator, which is a vector field. It includes three subcases, called the ``parallel,'' ``orthogonal'' and ``tilted'' cases. They are distinguished by the relation between the generator and a timelike vector field, which is identified as the fluid flow, if it exists. The ``tilted'' case, in which the generator is neither parallel nor orthogonal to the fluid flow, is the most general case among them. 

Self-similarity in general relativity was first defined by Cahill and Taub as the existence of a generator called a homothetic Killing vector field.~\cite{ct1971} Such self-similarity is often called homothety. For the perfect-fluid case, homothety restricts a barotropic equation of state to take the linear form $p=K\mu$, where $p$ and $\mu$ are the pressure and energy density, respectively, and $K$ is a constant. The dominant energy condition requires $-1 \le K \le 1$. Spherically symmetric homothetic solutions in the tilted case have been intensively investigated in various situations. It has been found both in Newtonian gravity and general relativity that a homothetic solution is an attractor solution in gravitational collapse~\cite{larson1969,penston1969,fc1993,hn1997,ti1999,mh2001,hms2003,op1987,op1988,op1990,harada1998,hm2001,harada2001} or a critical solution associated with critical phenomena.~\cite{choptuik1993,ec1994,kha,nc2000,hm2001,bcgn2002} Spherically symmetric homothetic solutions have been used in modeling the growth of primordial black holes~\cite{ch1974,bh1978} and the evolution of a cosmic void.~\cite{tomitavoid} Those in the tilted case were classified for the dust ($K=0$) case by Carr~\cite{carr2000} and for the perfect-fluid case by Carr and Coley.~\cite{cc2000} (See also Refs.~\citen{gnu1998,ccgnu2000,ccgnu2001}.) Coley showed that the Friedmann-Robertson-Walker (FRW) solution is the only spherically symmetric homothetic perfect-fluid solution in the parallel case.~\cite{coley1991} McIntosh showed that a stiff fluid ($K=1$) is the only compatible perfect fluid with homothety in the orthogonal case.~\cite{mcintosh1975} 
 
One possible type of self-similarity compatible with more general equations of state is kinematic self-similarity. It is one natural generalization of homothety and is defined by a generator called a kinematic self-similarity vector field.~\cite{ch1989,ch1991} Kinematic self-similarity is classified with respect to an index into four kinds, the first, second, zeroth and infinite kinds. In the context of kinematic self-similarity, homothety is considered the first kind. Thus, when we also take the relation between a self-similarity generator and a fluid flow into consideration, there are twelve subcases of kinematic self-similarity. Although there are some works on kinematic self-similar solutions,~\cite{coley1997,bc1998,sbc2001,blvw2002} the importance of kinematic self-similarity is still less clear than that of homothetic solutions. In a previous work by the present authors~\cite{mhio2002b} (I), it is assumed that the equation of state is polytropic or $p=K\mu$ and a classification of spherically symmetric kinematic self-similar solutions was presented.~\cite{mhio2002a,mhio2002b} In that work, we only treated the cases of a non-zero pressure and vacuum, leaving out the pressureless (i.e. the dust) case. Spherically symmetric dust solutions can be obtained in closed form with two arbitrary functions and are known as the Lema{\^ i}tre-Tolman-Bondi family of solutions.~\cite{LTB} Among them, spherically symmetric homothetic dust solutions have been used in analyzing naked singularities,~\cite{zannias1991,hp1991,lake1992} in modeling the growth of primordial black holes~\cite{ch1974} and in describing the evolution of a cosmic void.~\cite{tomitavoid} Spherically symmetric kinematic self-similar dust solutions could have as many important applications as homothetic solutions. In this paper, we extend I, add some corrections and complements, and complete the classification.

The organization of this paper is the following. In \S \ref{sec:sssandkss}, the basic equations in a spherically symmetric spacetime are presented, and kinematic self-similarity is briefly reviewed. In \S \ref{sec:dust}, we classify the spherically symmetric kinematic self-similar dust solutions of the second, zeroth and infinite kinds. In \S \ref{sec:correction}, we give corrections and complements to I.~\cite{mhio2002b} In \S \ref{sec:perfectfluid}, we classify the spherically symmetric homothetic perfect-fluid solutions together with vacuum solutions for which a homothetic Killing vector field is parallel or orthogonal to the fluid flow. Section \ref{sec:summary} is devoted to summary and discussion. The flatness of 3-surfaces for kinematic self-similar dust solutions are presented in Appendix~\ref{intcond}. We adopt units for which $c=1$.

\section{Spherically symmetric spacetime and kinematic self-similarity}
\label{sec:sssandkss}
The line element in a spherically symmetric spacetime is given by 
\begin{eqnarray}
ds^2 = -e^{2\Phi(t,r)}dt^2+e^{2\Psi(t,r)}dr^2+R(t,r)^2 d\Omega^2,
\end{eqnarray}
where $d\Omega^2=d\theta^2+\sin^2 \theta d\varphi^2$. We consider a perfect fluid as a matter field,
\begin{eqnarray}
T_{\mu\nu} &=& p(t,r)g_{\mu\nu}+[\mu(t,r)+p(t,r)]U_{\mu} U_{\nu}, \\
U_{\mu} &=& (-e^{\Phi},0,0,0),
\end{eqnarray}
where $U^{\mu}$ is the four-velocity of the fluid element. Here we have adopted comoving coordinates. Then, the Einstein equations and the equations of motion for the perfect fluid are reduced to the following simple form:
\begin{eqnarray}
\Phi_r &=& -\frac{p_r}{\mu+p}, \label{basic1}\\
\Psi_t &=& -\frac{\mu_t}{\mu+p}-\frac{2R_t}{R}, \label{basic2}\\
m_r &=& 4\pi \mu R_r R^2,   \label{basic3}\\
m_t &=& -4\pi p R_t R^2, \label{basic4}\\ 
0&=&-R_{tr}+\Phi_r R_{t}+\Psi_t R_r,\label{basic5}\\
m &=& \frac{1}{2G} R(1+e^{-2\Phi}{R_t}^2-e^{-2\Psi}
{R_r}^2 ),\label{basic6}
\end{eqnarray}
where the subscripts $t$ and $r$ indicate differentiation with respect to $t$ and $r$, and $m(t,r)$ is the so-called Misner-Sharp mass. The auxiliary equations from the Einstein equations are as follows:
\begin{eqnarray}
&&-\frac{e^{2\Phi}}{R^2}-\left[\left(\frac{R_t}{R}\right)^{2}+
2\frac{R_t}{R} \Psi_t \right] +e^{2\Phi-2\Psi}\left[2\frac{R_{rr}}{R}-
2\frac{R_r}{R} \Psi_r+\left(\frac{R_r}{R}\right)^{2}\right]\nonumber \\
&=&-8\pi G\mu e^{2\Phi},\label{00}\\
&&\frac{e^{2\Psi}}{R^2}+e^{2\Psi-2\Phi}\left[2\frac{R_{tt}}{R}-
2\frac{R_t}{R} \Phi_t+\left(\frac{R_t}{R}\right)^{2}\right] -\left[\left(\frac{R_r}{R}\right)^{2}+2\frac{R_r}{R} \Phi_r\right]\nonumber \\
&=& -8\pi G p e^{2\Psi},\label{11}
\end{eqnarray}
Five of the above eight equations are independent. 

On a hypersurface of constant $t$, the two-sphere $R=2Gm$ is a marginally trapped surface (or an apparent horizon). The region $R<2Gm$ is trapped, while the region $R>2Gm$ is not trapped. If we assume $R_r>0$, outgoing (ingoing) null geodesics satisfy
\begin{eqnarray}
\frac{dt}{dr}\big{|}_{\pm}=\pm e^{\Psi-\Phi},
\end{eqnarray}
where the subscripts $+$ and $-$ denote outgoing and ingoing, respectively. Along the trajectory of an apparent horizon, the relation
\begin{eqnarray}
\left(\frac{ds}{dr}\big{|}_{\mbox{AH}}\right)^2&=&-e^{2\Phi}\left[\left(\frac{dt}{dr}\big{|}_{\mbox{AH}}\right)^2-e^{2\Psi-2\Phi}\right],\\
&=&-e^{2\Phi}\left(\frac{dt}{dr}\big{|}_{\mbox{AH}}-\frac{dt}{dr}\big{|}_{+}\right)\left(\frac{dt}{dr}\big{|}_{\mbox{AH}}-\frac{dt}{dr}\big{|}_{-}\right) \label{trajectory}
\end{eqnarray}
is satisfied, where the subscript AH indicates that the derivative is taken at an apparent horizon. It is found from Eq.~(\ref{trajectory}) that the trajectory of the apparent horizon is spacelike when 
\begin{eqnarray}
\frac{dt}{dr}\big{|}_{-}<\frac{dt}{dr}\big{|}_{\mbox{AH}}<\frac{dt}{dr}\big{|}_{+}
\end{eqnarray}
is satisfied on it, while it is timelike when
\begin{eqnarray}
\frac{dt}{dr}\big{|}_{\mbox{AH}}<\frac{dt}{dr}\big{|}_{-}~~\mbox{or}~~\frac{dt}{dr}\big{|}_{+}<\frac{dt}{dr}\big{|}_{\mbox{AH}}
\end{eqnarray}
is satisfied on it. The trajectory of an apparent horizon is an outgoing (ingoing) null when 
\begin{eqnarray}
\frac{dt}{dr}\big{|}_{\mbox{AH}}=\frac{dt}{dr}\big{|}_{\pm}
\end{eqnarray}
is satisfied on it. 

A kinematic self-similarity vector ${\bf\xi}$ satisfies the conditions
\begin{eqnarray}
{\cal{L}}_{\bf\xi} h_{\mu\nu} &=&2\delta h_{\mu\nu},\label{kss}\\
{\cal{L}}_{\bf\xi} U_{\mu} &=&\alpha U_{\mu},\label{gss}
\end{eqnarray}
where $h_{\mu\nu} =g_{\mu\nu}+U_{\mu}U_{\nu}$ is the projection tensor, ${\cal{L}}_{\bf\xi}$ denotes Lie differentiation along ${\bf\xi}$, and $\alpha$ and $\delta$ are constants.~\cite{ch1989,ch1991,coley1997} The similarity transformation is characterized by the scale-independent ratio $\alpha/\delta$, which is referred to as the similarity index. 

In the case $\delta \ne 0$, $\delta$ can be set to unity, and the kinematic self-similarity vector ${\bf\xi}$ can be written
\begin{eqnarray}
\xi^{\mu}\frac{\partial}{\partial x^{\mu}}=(\alpha t+\beta) \frac{\partial}{\partial t}+r \frac{\partial}{\partial r},
\end{eqnarray}
if it is tilted, i.e. neither parallel nor orthogonal to the fluid flow. In the case $\alpha=1$, which corresponds to self-similarity of the first kind (for which $\beta$ can be set to zero), it follows that ${\bf\xi}$ is a homothetic vector, and the self-similarity variable $\xi$ is given by $r/t$ . In the case $\alpha=0$, which corresponds to self-similarity of the zeroth kind (for which $\beta$ can be rescaled to unity), the self-similarity variable is given by 
\begin{eqnarray}
\xi=r e^{-t}.
\end{eqnarray}
In the case $\alpha \ne 0, 1$, which corresponds to self-similarity of the second kind (for which $\beta$ can be set to zero), the self-similarity variable is given by 
\begin{eqnarray}
\xi=\frac{r}{(\alpha t)^{1/\alpha}}.
\end{eqnarray}
In the case $\delta \ne 0$, self-similarity implies that the metric functions can be written
\begin{eqnarray}
R=r S(\xi), \quad \Phi={\bar \Phi}(\xi), \quad \Psi={\bar \Psi}(\xi).\label{finitessform}
\end{eqnarray}
In the special case that $\delta=0$ and $\alpha \ne 0$, the self-similarity is referred to as that of the infinite kind (here, $\alpha=1$ is possible). The kinematic self-similarity vector ${\bf\xi}$ can be written
\begin{eqnarray}
\xi^{\mu}\frac{\partial}{\partial x^{\mu}}=t \frac{\partial}{\partial t}+r \frac{\partial}{\partial r}.
\end{eqnarray}
In this case, the self-similarity variable is given by 
\begin{eqnarray}
\xi=\frac{r}{t}.
\end{eqnarray}
In the case $\delta = 0$, self-similarity implies that the metric functions can be written
\begin{eqnarray}
R=S(\xi), \quad e^{\Phi}=e^{{\bar \Phi}(\xi)}, \quad e^{\Psi}=e^{{\bar \Psi}(\xi)}/r.
\end{eqnarray}
It is noted that ${\bf \xi}$ is a Killing vector in the case that $\delta=0$ and $\alpha = 0$. We omit the bars on ${\bar \Phi}$ and ${\bar \Psi}$ in \S \ref{sec:dusttilted} for simplicity. 

If the kinematic self-similarity vector ${\bf \xi}$ is parallel to the fluid flow, it can be written
\begin{eqnarray}
\xi^{\mu}\frac{\partial}{\partial x^{\mu}}=f(t) \frac{\partial}{\partial t},\label{xiparallel}
\end{eqnarray}
where $f(t)$ is an arbitrary function, and the self-similarity variable is now $r$. When we set $f(t)=t$, the metric can be written 
\begin{eqnarray}
ds^2 = -t^{2(\alpha-1)}e^{2{\bar \Phi}(r)}dt^2+t^2dr^2+t^2 S(r)^2 d\Omega^2 \label{parafinite}
\end{eqnarray}
for the first, second and zeroth kinds, and 
\begin{eqnarray}
ds^2 = -e^{2{\bar \Phi}(r)}dt^2+dr^2+S(r)^2 d\Omega^2 \label{parainfinite}
\end{eqnarray}
for the infinite kind. We omit the bars on ${\bar \Phi}$ in \S \ref{sec:dustparallel} and \S \ref{sec:perfectfluidparallel} for simplicity. 

If the kinematic self-similarity vector ${\bf \xi}$ is orthogonal to the fluid flow, we have
\begin{eqnarray}
\xi^{\mu}\frac{\partial}{\partial x^{\mu}}=g(r) \frac{\partial}{\partial r},\label{xiorthogonal}
\end{eqnarray}
where $g(r)$ is an arbitrary function, and the self-similarity variable is then $t$. When we set $g(r)=r$, the metric can be written 
\begin{eqnarray}
ds^2 = -r^{2\alpha}dt^2+e^{2{\bar \Psi}(t)}dr^2+r^2 S(t)^2 d\Omega^2 \label{orthfinite}
\end{eqnarray}
for the first, second and zeroth kinds, and
\begin{eqnarray}
ds^2 = -r^2dt^2+\frac{e^{2{\bar \Psi}(t)}}{r^2}dr^2+S(t)^2 d\Omega^2 \label{orthinfinite}
\end{eqnarray}
for the infinite kind. We omit the bars on ${\bar \Psi}$ in \S \ref{sec:dustorthogonal} and \S \ref{sec:perfectfluidorthogonal} for simplicity.

\section{Spherically symmetric kinematic self-similar dust solutions}
\label{sec:dust}
Setting $P_1=P_2=0$ in the basic equations for a perfect fluid given in I,~\cite{mhio2002b} we obtain those for a dust fluid. Since we classified the vacuum case in I,~\cite{mhio2002b} we omit the vacuum case in this section.

\subsection{Tilted case}
\label{sec:dusttilted}
In the tilted case for dust solutions of the second, zeroth and infinite kinds, $\Phi'=0$ is obtained from basic equations. Therefore we set $\exp(\Phi)=c_0$, where $c_0$ is a positive constant, in this subsection.

\subsubsection{Self-similarity of the second kind}
In the case of self-similarity of the second kind, the Einstein equations imply that the quantities $m$ and $\mu$ must be of the forms 
\begin{eqnarray}
2Gm&=&r\left[M_1(\xi)+\frac{r^2}{t^2}M_2(\xi)\right],\label{2ndm}\\
8\pi G \mu&=&\frac{1}{r^2}\left[W_1(\xi)+\frac{r^2}{t^2}W_2(\xi)\right],\label{2ndmu}
\end{eqnarray}
where $\xi=r/(\alpha t)^{1/\alpha}$. The basic equations are the following:
\begin{eqnarray}
M_1+M_1'&=&W_1S^2(S+S'),\label{2nd1}\\
3M_2+M_2'&=&W_2S^2(S+S'),\label{2nd2}\\
M_1'&=&0,\label{2nd3}\\
2\alpha M_2+M_2'&=&0,\label{2nd4}\\
M_1&=&S[1-e^{-2\Psi}(S+S')^2],\label{2nd5}\\
\alpha^2 M_2&=&SS'^2 c_0^{-2},\label{2nd6}\\
W_1'S&=&-W_1(\Psi'S+2S'),\label{2nd9}\\
(2\alpha W_2+W_2')S&=&-W_2(\Psi'S+2S'),\label{2nd10}\\
S''+S'&=&(S+S')\Psi',\label{2nd11}\\
S'(S'+2\Psi'S)&=&\alpha^2W_2 S^2 c_0^2, \label{2nd00a}\\
2S(S''+2S')-2\Psi'S(S+S')&=&-S'^2-S^2+e^{2\Psi}(1-W_1S^2),\label{2nd00b}\\
2S(S''+\alpha S')+S'^2&=&0,\label{2nd11a}\\
(S+S')^2&=&e^{2\Psi},\label{2nd11b}
\end{eqnarray}
where the prime denotes differentiation with respect to $\ln \xi$. Here we have implicitly assumed $\alpha t>0$. When we define $\xi \equiv r/(-\alpha t)^{1/\alpha}$ for $\alpha t<0$, basic equations identical to Eqs. (\ref{2nd1})--(\ref{2nd11b}) are obtained. We extend solutions analytically to the region in which $\alpha t<0$ if they are analytic at $t=0$. Such extended solutions can be obtained when we allow solutions to be defined for $\alpha t<0$. 

Equations (\ref{2nd11b}) and (\ref{2nd5}) give $M_1=0$, which implies $W_1=0$ from Eq. (\ref{2nd1}), because $|S+S'|=\exp(\Psi)$ cannot be zero. Equation (\ref{2nd11a}) gives the evolution equation for $S$,
\begin{eqnarray}
2y'+3y^2+2\alpha y=0,
\end{eqnarray}
where $y \equiv S'/S$. This equation can be integrated to yield
\begin{eqnarray}
S=|s_1-s_0 \xi^{-\alpha}|^{2/3},
\end{eqnarray}
where $s_0$ and $s_1$ are integration constants. In the present case, $s_0 \ne 0$ must be satisfied, because $s_0=0$ gives a vacuum spacetime. The relation $M_2=SS'^2/(\alpha^2 c_0^2)$ is obtained from Eq. (\ref{2nd6}), while $W_2=(3-2\alpha)y^2 /[\alpha^2 c_0^2(1+y)]$ is obtained from Eqs. (\ref{2nd00a}), (\ref{2nd11}) and (\ref{2nd11a}). The resulting solution is 
\begin{eqnarray}
e^{2\Phi}&=&c_0^2,\\
e^{2\Psi}&=&\frac{[s_1+((2\alpha/3)-1)s_0 \xi^{-\alpha}]^2}{|s_1-s_0 \xi^{-\alpha}|^{2/3}} \nonumber \\
&=&\frac{[s_1r^{\alpha}+((2\alpha/3)-1)s_0 \alpha t]^2}{r^{4\alpha/3}|s_1r^{\alpha}-s_0 \alpha t|^{2/3}},\\
R&=&r|s_1-s_0 \xi^{-\alpha}|^{2/3} \nonumber \\
&=&r^{1-2\alpha/3}|s_1r^{\alpha}-s_0 \alpha t|^{2/3},\\
8\pi G\mu&=&\frac{4(3-2\alpha)s_0^2 \xi^{-2\alpha}}{9c_0^2t^2(s_1-s_0 \xi^{-\alpha})[s_1+((2\alpha/3)-1)s_0 \xi^{-\alpha}]} \nonumber \\
&=&\frac{4(3-2\alpha)s_0^2 \alpha^2}{9c_0^2(s_1r^{\alpha}-s_0 \alpha t)[s_1r^{\alpha}+((2\alpha/3)-1)s_0 \alpha t]},\\
2Gm&=&\frac{4s_0^2r^3}{9c_0^2t^2}\xi^{-2\alpha} \nonumber \\
&=&\frac{4s_0^2\alpha^2}{9c_0^2}r^{3-2\alpha}.
\end{eqnarray}
When the time-reversal solution is also considered, we can choose $s_0>0$ and $s_1 \ge 0$ without loss of generality. With the coordinate transformations $t \to \bar{t}=c_0t$ and $r \to \bar{r}=|s_0\alpha/c_0|^{2/3}r^{(3-2\alpha)/3}$, we obtain a one-parameter family of solutions parametrized by $\kappa \equiv (s_1 c_0/(s_0\alpha))|c_0/(s_0\alpha)|^{2\alpha/(3-2\alpha)}$ for each $\alpha$, 
\begin{eqnarray}
ds^2&=&-d\bar{t}^2+\frac{9(\kappa\bar{r}^{3\alpha/(3-2\alpha)}+(2\alpha/3-1)\bar{t})^2}{(3-2\alpha)^2|\kappa\bar{r}^{3\alpha/(3-2\alpha)}-\bar{t}|^{2/3}}d\bar{r}^2 \nonumber \\
&&~~~~~~~~~~~~~~~~~~~~~~~~+\bar{r}^2|\kappa\bar{r}^{3\alpha/(3-2\alpha)}-\bar{t}|^{4/3}d\Omega^2,\label{solsecond1}\\
8\pi G\mu&=&\frac{4(3-2\alpha)}{9(\kappa\bar{r}^{3\alpha/(3-2\alpha)}+(2\alpha/3-1)\bar{t})(\kappa\bar{r}^{3\alpha/(3-2\alpha)}-\bar{t})},\label{solsecond2}\\
2Gm&=&\frac49 \bar{r}^3. \label{solsecond3}
\end{eqnarray}
This solution was obtained previously by Carter and Henriksen~\cite{ch1989} and has been treated by several authors.~\cite{bc1998,blvw2002} This solution belongs to the Lema{\^ i}tre-Tolman-Bondi family of solutions for the marginally bound case. We call this solution the second-kind kinematic self-similar Lema{\^ i}tre-Tolman-Bondi (KSS-LTB) solution. We omit the bars on ${\bar t}$ and ${\bar r}$ hereafter for simplicity. If $s_1=0$ (i.e., $\kappa=0$), this solution gives the flat FRW solution for any $\alpha \ne 0, 1$. When $s_1 \ne 0$ (i.e., $\kappa \ne 0$), $\alpha \ne 3/2$ must be satisfied for the solution to be a non-vacuum solution. This solution with $\kappa \ne 0$ is asymptotically flat FRW solution for $r \to \infty$ with $\alpha<0$ or $3/2<\alpha$ and for $r \to 0$ with $0<\alpha<3/2~(\alpha \ne 1)$. It is noted that for $\kappa \ne 0$, $r \to +0$ corresponds to $R \to +\infty$ for $\alpha>3/2$ and to $R \to +0$ otherwise. The energy density is positive in the regions satisfying 
\begin{equation}
t<\kappa r^{3\alpha/(3-2\alpha)},~~[3/(3-2\alpha)]\kappa r^{3\alpha/(3-2\alpha)}<t~~\mbox{for}~~0<\alpha<3/2,~\alpha \ne 1
\end{equation}
and 
\begin{equation}
t<[3/(3-2\alpha)]\kappa r^{3\alpha/(3-2\alpha)},~~\kappa r^{3\alpha/(3-2\alpha)}<t~~\mbox{for}~~\alpha<0,~3/2<\alpha.
\end{equation}
The singular surfaces are $t=\kappa r^{3\alpha/(3-2\alpha)}$ and $t=[3/(3-2\alpha)]\kappa r^{3\alpha/(3-2\alpha)}$. The former is a shell-focusing singularity, because the circumferential radius $R$ is zero there, while the latter is a shell-crossing singularity, because the quantity 
\begin{eqnarray}
\frac{\partial R}{\partial r}=
\left\{
\begin{array}{lr}
\{(3/(3-2\alpha))\kappa r^{3\alpha/(3-2\alpha)}-t\}/(\kappa r^{3\alpha/(3-2\alpha)}-t)^{1/3}  \\
~~~~~~~~~~~~~~~~~~~~~~~~~~~~~~~~~~~~~~~~~~~~~~\mbox{for}~~t<\kappa r^{3\alpha/(3-2\alpha)} ,\\
\{t-(3/(3-2\alpha))\kappa r^{3\alpha/(3-2\alpha)}\}/(t-\kappa r^{3\alpha/(3-2\alpha)})^{1/3}  \\
~~~~~~~~~~~~~~~~~~~~~~~~~~~~~~~~~~~~~~~~~~~~~~\mbox{for}~~t>\kappa r^{3\alpha/(3-2\alpha)}
\end{array}\right.
\end{eqnarray}
is zero there. For the trajectory of the shell-crossing singularity, the relation 
\begin{eqnarray}
R=\big{|}\frac{2\alpha}{3-2\alpha}\kappa\big{|}^{2/3}r^{3/(3-2\alpha)}
\end{eqnarray}
is satisfied. The fluid is collapsing in the region satisfying $t<\kappa r^{3\alpha/(3-2\alpha)}$ and expanding in the region satisfying $t>\kappa r^{3\alpha/(3-2\alpha)}$, because the quantity
\begin{eqnarray}
\frac{\partial R}{\partial t}=-\mbox{sign}(\kappa r^{3\alpha/(3-2\alpha)}-t)\frac23r|\kappa r^{3\alpha/(3-2\alpha)}-t|^{-1/3}
\end{eqnarray}
is negative in the former case and positive in the latter case. The trapped region is given by 
\begin{eqnarray}
\kappa r^{3\alpha/(3-2\alpha)}-(8/27)r^3<t<\kappa r^{3\alpha/(3-2\alpha)}
\end{eqnarray}
and 
\begin{eqnarray}
\kappa r^{3\alpha/(3-2\alpha)}<t<\kappa r^{3\alpha/(3-2\alpha)}+(8/27)r^3.
\end{eqnarray}
The apparent horizons are 
\begin{eqnarray}
t=\kappa r^{3\alpha/(3-2\alpha)}+(8/27)r^3,~~(\mbox{AH1}) \label{AHsecond1}
\end{eqnarray}
and
\begin{eqnarray}
t=\kappa r^{3\alpha/(3-2\alpha)}-(8/27)r^3.~~(\mbox{AH2}) \label{AHsecond2}
\end{eqnarray}
Here we consider the trajectories of the apparent horizons. We consider only the regions with $\mu>0$ in which $R_r>0$ are satisfied. From Eqs (\ref{AHsecond1}) and (\ref{AHsecond2}), we obtain 
\begin{eqnarray}
\frac{dt}{dr}\big{|}_{\mbox{AH}}=
\left\{
\begin{array}{ll}
\{3\alpha\kappa/(3-2\alpha)\} r^{(5\alpha-3)/(3-2\alpha)}+ (8/9)r^2 \quad \mbox{for AH1},\\
\{3\alpha\kappa/(3-2\alpha)\} r^{(5\alpha-3)/(3-2\alpha)}- (8/9)r^2 \quad \mbox{for AH2}.\\
\end{array}\right.
\end{eqnarray}
AH1 has its extremum at $r=r_1 \equiv [27\alpha\kappa/\{8(2\alpha-3)\}]^{(3-2\alpha)/\{9(1-\alpha)\}}$ for $3/2<\alpha$ or $\alpha<0$, while does not for $0<\alpha<3/2~(\alpha \ne 1)$. AH2 has its extremum at $r=r_2 \equiv [27\alpha\kappa/\{8(3-2\alpha)\}]^{(3-2\alpha)/\{9(1-\alpha)\}}$ for $0<\alpha<3/2~(\alpha \ne 1)$, while does not for $3/2<\alpha$ or $\alpha<0$. For the null geodesics on the apparent horizon, the relation
\begin{eqnarray}
\frac{dt}{dr}\big{|}_{\pm}=
\left\{
\begin{array}{ll}
\pm|\{3\alpha\kappa/(3-2\alpha)\}r^{(5\alpha-3)/(3-2\alpha)} - (4/9)r^2|\\
~~~~~~~~~~~~~~~~~~~~~~~~~~~~~~~~~~~~~~~~~~~~~~~~~~~~~~~\mbox{for AH1},\\
\pm|\{3\alpha\kappa/(3-2\alpha)\}r^{(5\alpha-3)/(3-2\alpha)} + (4/9)r^2|\\
~~~~~~~~~~~~~~~~~~~~~~~~~~~~~~~~~~~~~~~~~~~~~~~~~~~~~~~\mbox{for AH2}\\
\end{array}\right.
\end{eqnarray}
are satisfied. First we consider $r>0$. For $\alpha<0,~3/2<\alpha$, the trajectory of AH1 is spacelike (timelike) for 
\begin{eqnarray}
r^{9(1-\alpha)/(3-2\alpha)}<(>)-27\alpha\kappa/[2(3-2\alpha)], 
\end{eqnarray}
while it is outgoing null for 
\begin{eqnarray}
r^{9(1-\alpha)/(3-2\alpha)}=-27\alpha\kappa/[2(3-2\alpha)] \equiv r_3^{9(1-\alpha)/(3-2\alpha)}. 
\end{eqnarray}
For $0<\alpha<3/2~(\alpha \ne 1)$, the trajectory of AH1 is timelike. For $\alpha<0,~3/2<\alpha$, the trajectory of AH2 is timelike. For $0<\alpha<3/2$ and $\alpha \ne 1$, the trajectory of AH2 is spacelike (timelike) for 
\begin{eqnarray}
r^{9(1-\alpha)/(3-2\alpha)}<(>)27\alpha\kappa/[2(3-2\alpha)],
\end{eqnarray}
while it is ingoing null for 
\begin{eqnarray}
r^{9(1-\alpha)/(3-2\alpha)}=27\alpha\kappa/[2(3-2\alpha)] \equiv r_4^{9(1-\alpha)/(3-2\alpha)}.
\end{eqnarray}
Next we consider $r=0$. For both AH1 and AH2, the relation
\begin{eqnarray}
\lim_{r \to 0}\left(\frac{dt}{dr}\big{|}_{\mbox{AH}}\big{/}\frac{dt}{dr}\big{|}_{\pm}\right)=\pm \mbox{sign}(\alpha/(3-2\alpha)) \label{ratiosecond}
\end{eqnarray}
is satisfied for $\alpha<1$ and $3/2<\alpha$, so that both AH1 and AH2 are ingoing (outgoing) null at $r=0$ for $\alpha<0$ and $3/2<\alpha$ ($0<\alpha<1$). For $1<\alpha<3/2$, the left-hand-side of equation (\ref{ratiosecond}) is equal to $\pm 2$ and $\mp 2$ for AH1 and AH2, respectively, so that they are not null at $r=0$, and the results for $r>0$ can be applied for $r=0$.

\begin{figure}[htb]
\centerline{
\epsfxsize 11cm
\epsfbox{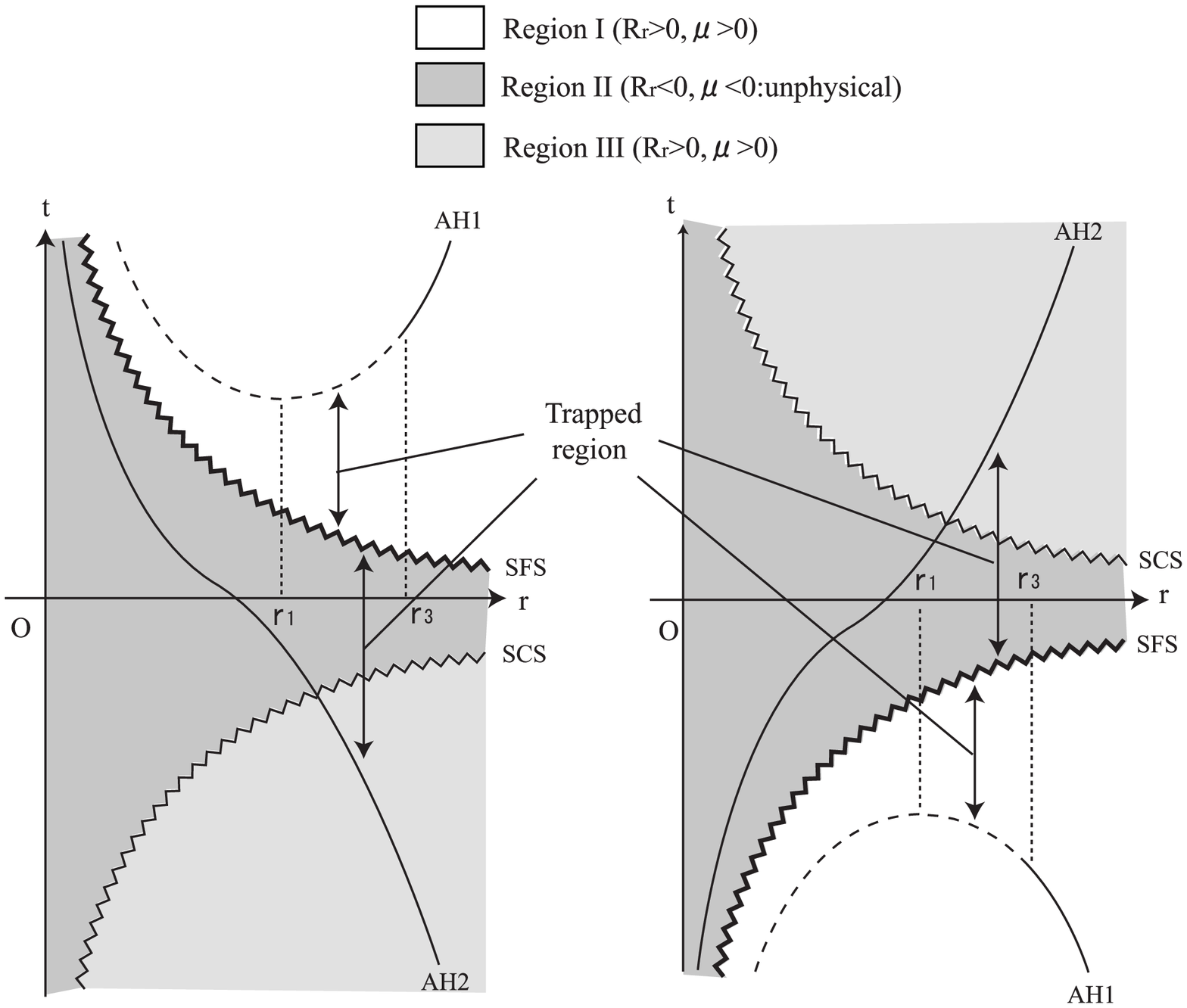}}
\vspace{0cm}
\caption{The figure on the left is for the second-kind KSS-LTB solution [Eqs.~(\ref{solsecond1})--(\ref{solsecond3})] with $\kappa \ne 0$ for $\alpha>3/2$, and that on the right is for its time time-reversal solution. AH, SFS and SCS indicate an apparent horizon, shell-focusing singularity and shell-crossing singularity, respectively. Each diagram is divided by SFS and SCS into three regions, which we call regions I, II and III. Region II is unphysical, because $\mu<0$ there. A solid curve represents an apparent horizon whose trajectory is timelike, while a dashed curve represents an apparent horizon whose trajectory is spacelike. $r=r_1$ corresponds to the extremum of AH1. The trajectory of AH1 is outgoing (ingoing) null at $r=r_3$ in the left (right) figure. It is noted that $r \to +0$ corresponds to $R \to +\infty$ for this $\alpha$.}
\label{second1}
\end{figure}
\begin{figure}[htb]
\centerline{
\epsfxsize 11cm
\epsfbox{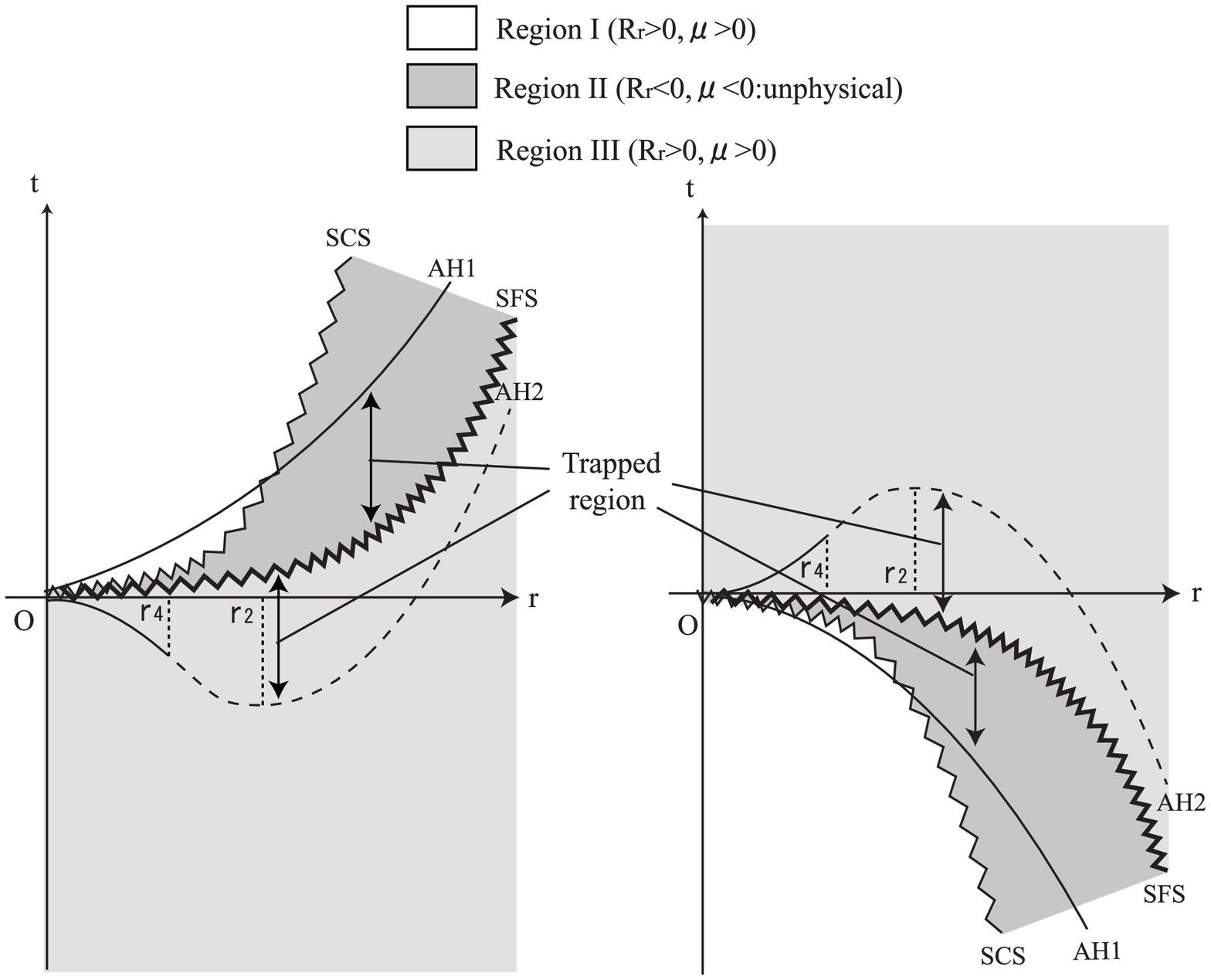}}
\vspace{0cm}
\caption{Same as Fig.~\ref{second1}, but for $1<\alpha<3/2$. Here, $r=r_2$ corresponds to the extremum of AH2. The trajectory of AH2 is ingoing (outgoing) null at $r=r_4$ in the left (right) figure.}
\label{second2b}
\end{figure}
\begin{figure}[htb]
\centerline{
\epsfxsize 11cm
\epsfbox{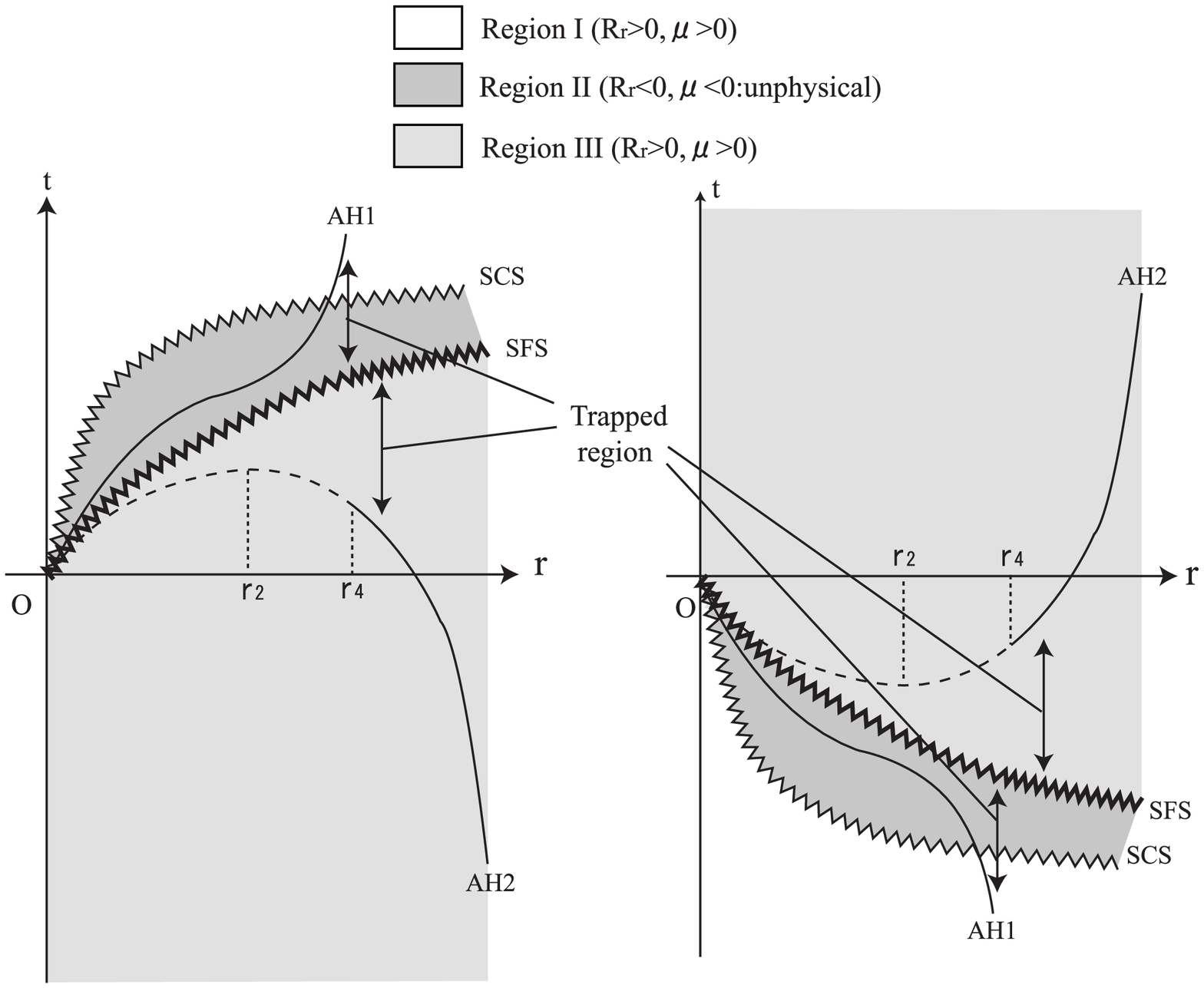}}
\vspace{0cm}
\caption{Same as Fig.~\ref{second1}, but for $0<\alpha<1$. Here, $r=r_2$ corresponds to the extremum of AH2. The trajectory of AH2 is ingoing (outgoing) null at $r=r_4$ in the left (right) figure.}
\label{second2a}
\end{figure}
\begin{figure}[htb]
\centerline{
\epsfxsize 11cm
\epsfbox{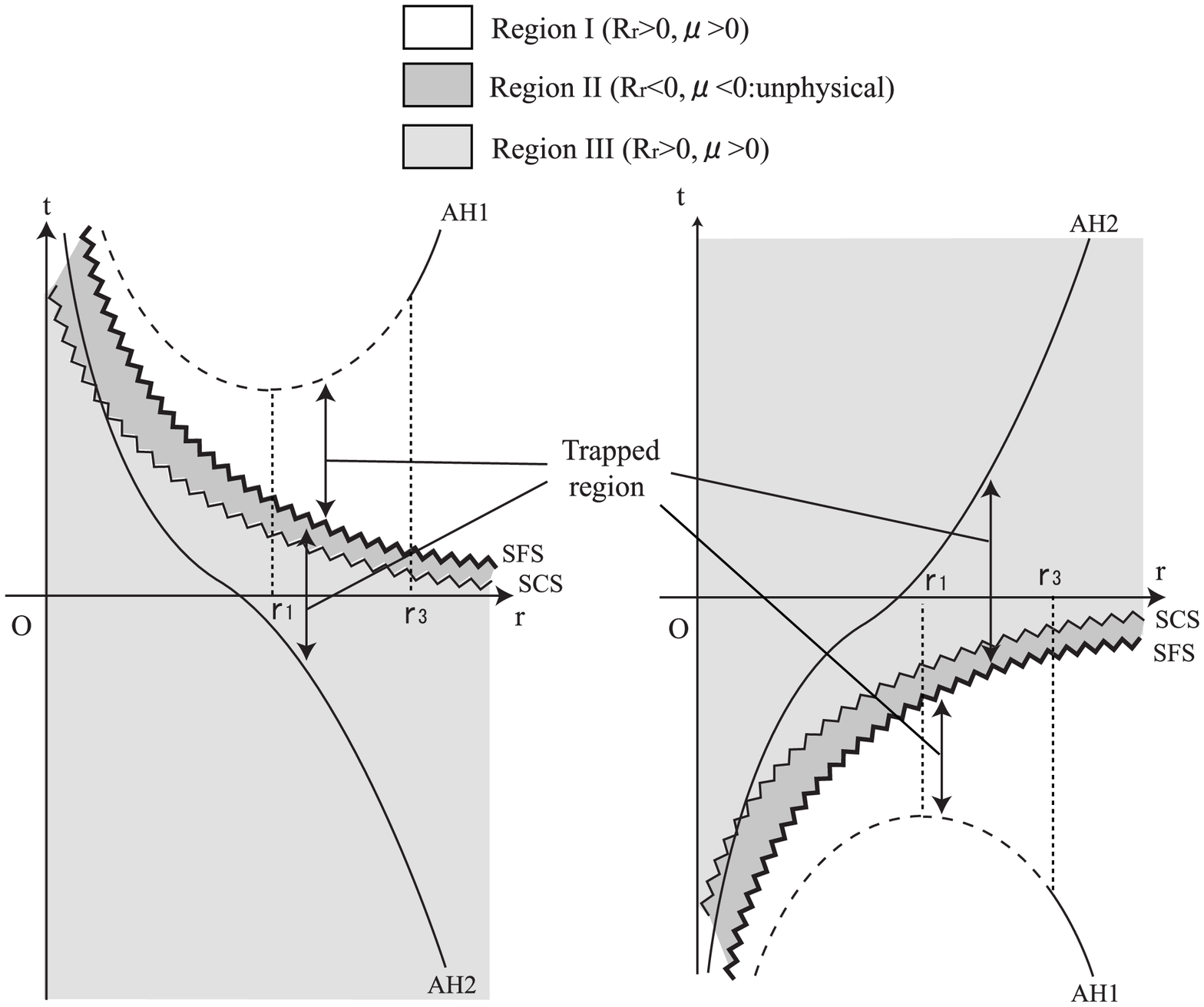}}
\vspace{0cm}
\caption{Same as Fig.~\ref{second1}, but for $\alpha<0$. $r=r_1$ corresponds to the extremum of AH1. The trajectory of AH1 is outgoing (ingoing) null at $r=r_3$ in the left (right) figure.}
\label{second3}
\end{figure}
Schematic figures for this solution are displayed in Figs.~\ref{second1}--\ref{second3}, along with those for its time-reversal solutions. For all $\alpha~(\ne 0,1$ or $3/2)$, the solution is unphysical in region II, in which the energy density is negative. We consider only regions with positive energy density. 

For $\alpha>3/2$, region I on the left in Fig.~\ref{second1} represents a decaying white hole, while region I in its time-reversal solution represents a growing black hole. Region III on the left in Fig.~\ref{second1} represents the outer region of a contracting shell-crossing singularity that starts from $R=+\infty$ at $t=-\infty$. The singularity reaches the physical center at $t=-0$. Region III in its time-reversal solution represents the outer region of an expanding shell-crossing singularity that starts from $R=+0$ at $t=+0$. 

For $0<\alpha<3/2$ and $\alpha \ne 1$, region I on the left in each of Figs.~\ref{second2b} and \ref{second2a} represents the inner region of an expanding shell-crossing singularity, which forms at the physical center at $t=+0$. Region I in each of the time-reversal solutions in Figs.~\ref{second2b} and \ref{second2a} represents the inner region of a contracting shell-crossing singularity that starts from $R=+\infty$ at $t=-\infty$. The singularity reaches the physical center at $t=-0$. Region III on the left in each of Figs.~\ref{second2b} and \ref{second2a} represents the formation of a shell-focusing singularity from a regular spacetime. Because the singularity is covered for $1<\alpha<3/2$, this solution represents the formation and growth of a black hole. For $0<\alpha<1$, further study is needed to determine whether this singularity is naked or covered. Region III on the right in each of Figs.~\ref{second2b} and \ref{second2a} represents a decaying white hole. The white hole singularity disappears at $t=0$ and leaves a regular spacetime. 

For $\alpha<0$, region I on the left in Fig.~\ref{second3} represents a decaying white hole, while region I in its time-reversal solution represents a growing black hole. Region III on the left in Fig.~\ref{second3} represents the formation of a non-central shell-crossing singularity from a regular spacetime. The singularity forms at $R=+\infty$ at $t=+0$ and contracts with time. For $t>0$, it represents the inner region of the singularity. Region III on the right in Fig.~\ref{second3} represents the inner region of an expanding  non-central shell-crossing singularity. This singularity reaches $R=+\infty$ at $t=-0$, then disappears and leaves a regular spacetime.

\subsubsection{Self-similarity of the zeroth kind}
In the case of self-similarity of the zeroth kind, the Einstein equations imply that the quantities $m$ and $\mu$ must be of the forms 
\begin{eqnarray}
2Gm&=&r[M_1(\xi)+r^2 M_2(\xi)],\label{zerom}\\
8\pi G \mu&=&\frac{1}{r^2}[W_1(\xi)+r^2W_2(\xi)],\label{zeromu}
\end{eqnarray}
where $\xi=r e^{-t}$. In this case, the basic equations are the following:
\begin{eqnarray}
M_1+M_1'&=&W_1S^2(S+S'),\label{zero1}\\
3M_2+M_2'&=&W_2S^2(S+S'),\label{zero2}\\
M_1'&=&0,\label{zero3}\\
M_2'&=&0,\label{zero4}\\
M_1&=&S[1-e^{-2\Psi}(S+S')^2],\label{zero5}\\
M_2&=&SS'^2 c_0^{-2},\label{zero6}\\
W_1'S&=&-W_1(\Psi'S+2S'),\label{zero9}\\
W_2'S&=&-W_2(\Psi'S+2S'),\label{zero10}\\
S''+S'&=&(S+S')\Psi',\label{zero11}\\
S'(S'+2\Psi'S)&=&W_2 S^2 c_0^2, \label{zero00a}\\
2S(S''+2S')-2\Psi'S(S+S')&=&-S'^2-S^2+e^{2\Psi}(1-W_1S^2),\label{zero00b}\\
2SS''+S'^2&=&0,\label{zero11a}\\
(S+S')^2&=&e^{2\Psi},\label{zero11b}
\end{eqnarray}
where the prime denotes the differentiation with respect to $\ln \xi$. Equation (\ref{zero11b}) gives $\exp(2\Psi)=(S+S')^2$, so that $M_1=0$ is concluded from Eq.~(\ref{zero5}). The equality $M_1=0$ implies $W_1=0$, from Eq. (\ref{zero1}), because $|S+S'|=\exp(\Psi)$ cannot be zero. Equation (\ref{zero11a}) gives the evolution equation for $S$,
\begin{eqnarray}
2y'+3y^2=0,
\end{eqnarray}
where $y \equiv S'/S$. This equation can be integrated to yield
\begin{eqnarray}
S=|s_1-s_0 \ln\xi|^{2/3},
\end{eqnarray}
where $s_0$ and $s_1$ are integration constants. The relation $s_0 \ne 0$ must be satisfied for the solution to be a non-vacuum solution. The equality $M_2=SS'^2/c_0^2$ is obtained from Eq. (\ref{zero6}), while $W_2=3y^2 /[c_0^2(1+y)]$ is obtained from Eqs. (\ref{zero00a}), (\ref{zero11}) and (\ref{zero11a}). The resulting solution is 
\begin{eqnarray}
e^{2\Phi}&=&c_0^2,\\
e^{2\Psi}&=&\frac{(s_1-2s_0/3-s_0\ln\xi)^2}{(s_1-s_0 \ln\xi)^{2/3}} \nonumber \\
&=&\frac{(s_1-2s_0/3-s_0(\ln r-t))^2}{|s_1-s_0(\ln r-t)|^{2/3}},\\
R&=&r|s_1-s_0 \ln\xi|^{2/3} \nonumber \\
&=&r|s_1-s_0(\ln r-t)|^{2/3},\\
8\pi G\mu&=&\frac{4s_0^2}{3c_0^2(s_1-s_0 \ln\xi)(s_1-2s_0/3-s_0\ln\xi)} \nonumber \\
&=&\frac{4s_0^2}{3c_0^2(s_1-s_0(\ln r-t))(s_1-2s_0/3-s_0(\ln r-t))},\\
2Gm&=&\frac{4s_0^2r^3}{9c_0^2}.
\end{eqnarray}
With the coordinate transformations $t \to \bar{t}=c_0t+c_0s_1/s_0+(2c_0/3)\ln(|s_0|/c_0)$ and $r \to \bar{r}=(|s_0|/c_0)^{2/3}r$, a one-parameter family of solutions parametrized by $c_0$, 
\begin{eqnarray}
ds^2&=&-d\bar{t}^2+\frac{(\bar{t}-2c_0/3-c_0\ln \bar{r})^2}{|\bar{t}-c_0\ln \bar{r}|^{2/3}}d\bar{r}^2+\bar{r}^2|\bar{t}-c_0\ln \bar{r}|^{4/3}d\Omega^2,\label{solzero1}\\
8\pi G\mu&=&\frac{4}{3(\bar{t}-c_0\ln \bar{r})(\bar{t}-2c_0/3-c_0\ln \bar{r})},\label{solzero2}\\
2Gm&=&\frac49 \bar{r}^3, \label{solzero3}
\end{eqnarray}
is obtained. This solution was obtained previously by Benoit and Coley.~\cite{bc1998} (See also Ref.~\citen{blvw2002}.) This solution belongs to the Lema{\^ i}tre-Tolman-Bondi family of solutions for the marginally bound case. We call this solution the zeroth-kind KSS-LTB solution. We omit the bars on ${\bar t}$ and ${\bar r}$ hereafter for simplicity. For this solution, the energy density is positive in the regions satisfying 
\begin{eqnarray}
t<c_0 \ln r, \quad c_0\ln r+2c_0/3<t.
\end{eqnarray}
The singular surfaces are $t=c_0\ln r$ and $t=c_0\ln r+2c_0/3$. The former is a shell-focusing singularity, because the circumferential radius $R$ is zero there, while the latter is a shell-crossing singularity, because the quantity
\begin{eqnarray}
\frac{\partial R}{\partial r}=
\left\{
\begin{array}{rl}
\{-t+c_0\ln r+(2c_0/3)\}/(-t+c_0\ln r)^{1/3}~~\mbox{for}~~t<c_0\ln r ,\\
\{t-c_0\ln r-(2c_0/3)\}/(t-c_0\ln r)^{1/3}~~\mbox{for}~~t>c_0\ln r,
\end{array}\right.
\end{eqnarray}
is zero there. The fluid is collapsing in the regions in which $t<c_0\ln r$ and expanding in the regions in which $t>c_0\ln r$, because the quantity 
\begin{eqnarray}
\frac{\partial R}{\partial t}=\mbox{sign}(t-c_0\ln r)\frac23r|t-c_0\ln r|^{-1/3}
\end{eqnarray}
is negative in the former case and positive in the latter case. The trapped region is given by 
\begin{eqnarray}
c_0\ln r-(8/27)r^3<t<c_0\ln r
\end{eqnarray}
and
\begin{eqnarray}
c_0\ln r<t<c_0\ln r+(8/27)r^3.
\end{eqnarray}
The apparent horizons are 
\begin{eqnarray}
t=c_0\ln r+(8/27)r^3,~~(\mbox{AH1}) \label{AHzeroth1}
\end{eqnarray}
and 
\begin{eqnarray}
t=c_0\ln r-(8/27)r^3.~~(\mbox{AH2}) \label{AHzeroth2}
\end{eqnarray}
Here we consider the trajectories of the apparent horizons. We consider only the solution with $\mu>0$ in which $R_r>0$ are satisfied. From Eqs.~(\ref{AHzeroth1}) and (\ref{AHzeroth2}), we obtain 
\begin{eqnarray}
\frac{dt}{dr}\big{|}_{\mbox{AH}}=
\left\{
\begin{array}{ll}
c_0/r + (8/9)r^2 \quad \mbox{for AH1},\\
c_0/r - (8/9)r^2 \quad \mbox{for AH2}.\\
\end{array}\right.
\end{eqnarray}
AH2 has its extremum at $r=r_5 \equiv (9c_0/8)^{1/3}$, while AH1 does not. For the null geodesics on the apparent horizon, the relation
\begin{eqnarray}
\frac{dt}{dr}\big{|}_{\pm}=
\left\{
\begin{array}{ll}
\pm |-c_0/r + (4/9)r^2|&\mbox{for AH1},\\
\pm |-c_0/r - (4/9)r^2|&\mbox{for AH2}\\
\end{array}\right.
\end{eqnarray}
are satisfied. For $r>0$, the trajectory of AH1 is timelike, while the trajectory of AH2 is spacelike (timelike) for 
\begin{eqnarray}
r^{3}<(>)(9/2)c_0.
\end{eqnarray}
The trajectory of AH2 is ingoing null for 
\begin{eqnarray}
r^{3}=(9/2)c_0 \equiv r_6^3.
\end{eqnarray}
Next, we consider $r=0$. The inequality $\mu<0$ is satisfied for $r \to 0$ along AH1, which is unphysical, and therefore we consider only AH2. For AH2, the relation
\begin{eqnarray}
\lim_{r \to 0}\left(\frac{dt}{dr}\big{|}_{\mbox{AH}}\big{/}\frac{dt}{dr}\big{|}_{\pm}\right)=\pm 1 \label{ratiozeroth}
\end{eqnarray}
is satisfied, so that AH2 is outgoing null at $r=0$.

\begin{figure}[htb]
\centerline{
\epsfxsize 11cm
\epsfbox{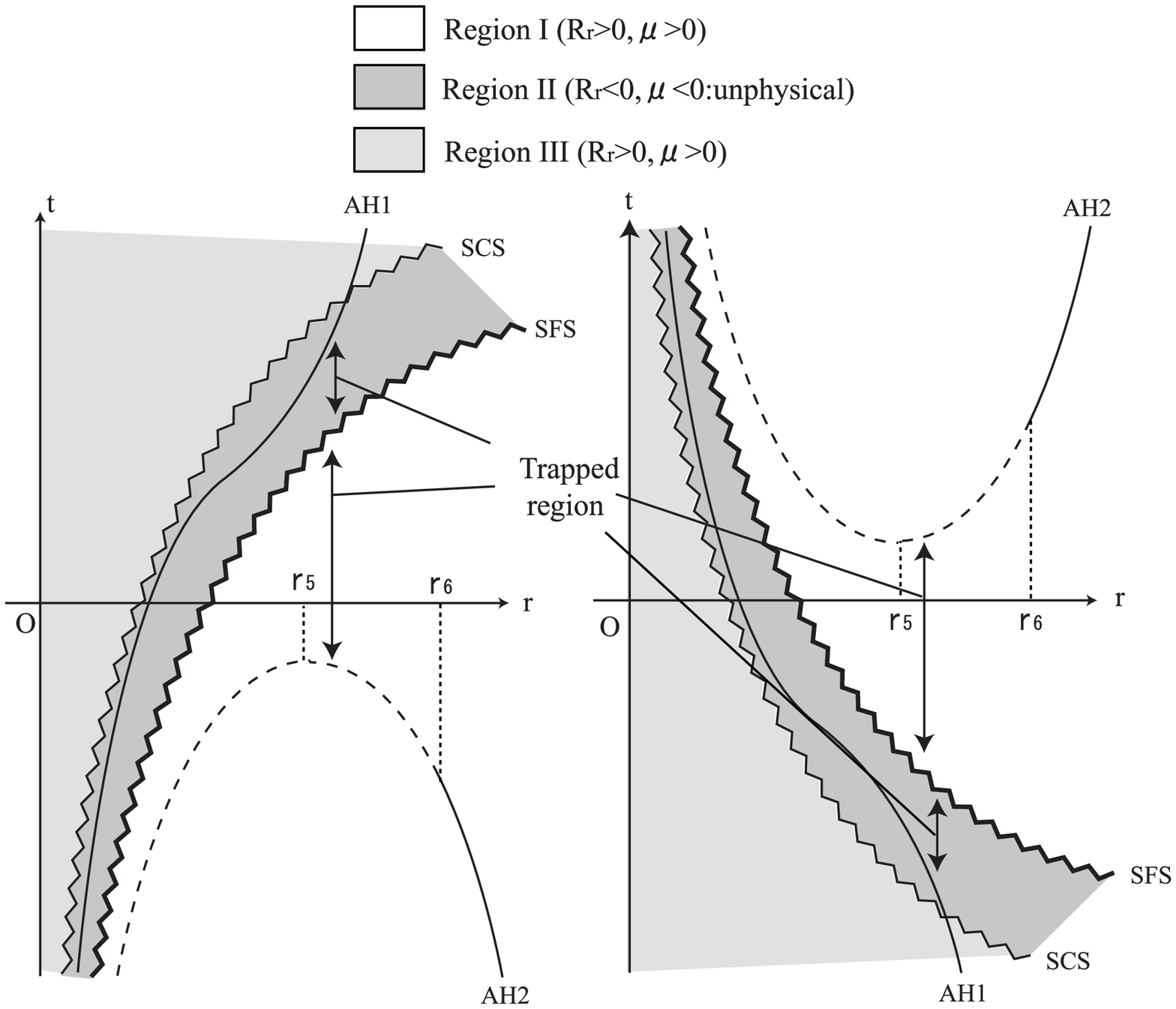}}
\vspace{0cm}
\caption{The figure on the left is for the zeroth-kind KSS-LTB solution [Eqs~(\ref{solzero1})--(\ref{solzero3})], and that on the right is for its time-reversal solution. AH, SFS and SCS indicate an apparent horizon, a shell-focusing singularity and a shell-crossing singularity, respectively. Each diagram is divided by SFS and SCS into three regions, which we call regions I, II and III. Region II is unphysical, because $\mu<0$ there. A solid curve represents an apparent horizon whose trajectory is timelike, while a dashed curve represents an apparent horizon whose trajectory is spacelike. $r=r_5$ corresponds to the extremum of AH2. The trajectory of AH2 is ingoing (outgoing) null at $r=r_6$ in the left (right) figure.}
\label{zero}
\end{figure}
A schematic figure for this solution is given in Fig.~\ref{zero}, along with that for its time-reversal solution. The solution is unphysical in region II, in which the energy density is negative. Region I on the left in Fig.~\ref{zero} represents a growing black hole, while that in its time-reversal solution represents a decaying white hole. Region III on the left in Fig.~\ref{zero} represents the inner region of an expanding non-central shell-crossing singularity that starts from $R=+0$ at $t=-\infty$, while that in its time-reversal solution represents the inner region of a contracting non-central shell-crossing singularity that starts from $R=+\infty$ at $t=-\infty$.

\subsubsection{Self-similarity of the infinite kind}
In the case of self-similarity of the infinite kind, the Einstein equations imply that the quantities $m$ and $\mu$ must be of the forms 
\begin{eqnarray}
2Gm&=&M_1(\xi)/t^2+M_2(\xi),\label{infm}\\
8\pi G \mu&=&W_1(\xi)/t^2+W_2(\xi),\label{infmu}
\end{eqnarray}
where $\xi=r/t$. In this case, the basic equations are the following:
\begin{eqnarray}
M_1'&=&W_1S^2 S',\label{inf1}\\
M_2'&=&W_2S^2 S'.\label{inf2}\\
2M_1+M_1'&=&0,\label{inf3}\\
M_2'&=&0,\label{inf4}\\
M_1&=&c_0^{-2}S S'^2,\label{inf5}\\
M_2&=&S(1-e^{-2\Psi}S'^2),\label{inf6}\\
(2W_1+W_1')S&=&-W_1(\Psi'S+2S'),\label{inf9}\\
W_2'S&=&-W_2(\Psi'S+2S'),\label{inf10}\\
S''&=&S'\Psi'\label{inf11},\\
W_1S^2c_0^2&=&S'(2\Psi'S+S'),\label{inf00a}\\
(1-W_2S^2)e^{2\Psi}&=&2SS''+S'^2-2\Psi'S'S,\label{inf00b}\\
e^{2\Psi}&=&S'^2,\label{inf11a}\\
0&=&2S''S+2SS'+S'^2,\label{inf11b}
\end{eqnarray}
where the prime denotes the differentiation with respect to $\ln \xi$. Here we have implicitly assumed $t>0$. When we define $\xi \equiv r/(-t)$ for $t<0$, equations identical to Eqs. (\ref{inf1})--(\ref{inf11b}) are obtained. We extend solutions analytically to the region in which $t<0$ if they are analytic at $t=0$. Such extended solutions can be obtained when we allow solutions to be defined for $t<0$. 

Equation (\ref{inf11a}) gives $\exp(2\Psi)=S'^2$, so that $M_2=0$ is concluded from Eq. (\ref{inf6}). The equality $M_2=0$ implies $W_2=0$, from Eq. (\ref{inf2}), because $|S'|=\exp(\Psi)$ cannot be zero. Equation (\ref{inf11b}) gives the evolution equation for $S$,
\begin{eqnarray}
2y'+3y^2+2 y=0,
\end{eqnarray}
where $y \equiv S'/S$. This equation can be integrated to yield
\begin{eqnarray}
S=|s_1-s_0 \xi^{-1}|^{2/3},
\end{eqnarray}
where $s_0$ and $s_1$ are integration constants. Here, $s_0$ cannot be zero, because this gives the contradictory relation $\exp(2\Psi)=0$. The relation $M_1=SS'^2/c_0^2$ is obtained from Eq. (\ref{inf5}), while $W_1=-2y/c_0^2$ is obtained from Eqs. (\ref{inf00a}), (\ref{inf11}) and (\ref{inf11b}). The resulting solution is
\begin{eqnarray}
e^{2\Phi}&=&c_0^2,\\
e^{2\Psi}/r^2&=&\frac49 s_0^2 r^{-2}\xi^{-2}|s_1-s_0 \xi^{-1}|^{-2/3} \nonumber \\
&=&\frac{4s_0^2t^2}{9r^{10/3}}|s_1r-s_0 t|^{-2/3},\\
R&=&S=|s_1-s_0 \xi^{-1}|^{2/3} \nonumber \\
&=&r^{-2/3}|s_1r-s_0t|^{2/3},\\
8\pi G\mu&=&\frac{-4s_0 \xi^{-1}}{3c_0^2 t^2(s_1-s_0 \xi^{-1})} \nonumber \\
&=&\frac{-4s_0}{3c_0^2 t(s_1r-s_0 t)},\\
2Gm&=&\frac{4s_0^2 \xi^{-2}}{9c_0^2 t^2},\nonumber \\
&=&\frac{4s_0^2}{9c_0^2 r^2}.
\end{eqnarray}
When the time-reversal solution is also considered, we can choose $s_0>0$ and $s_1 \ge 0$ without loss of generality. With the coordinate transformations $t \to \bar{t}=c_0t$ and $r \to \bar{r}=(s_0/c_0)^{2/3}r^{-2/3}$, a one-parameter family of solutions parametrized by $s_1$, 
\begin{eqnarray}
ds^2&=&-d\bar{t}^2+\bar{t}^{2}|s_1\bar{r}^{-3/2}-\bar{t}|^{-2/3}d\bar{r}^2+\bar{r}^2|s_1\bar{r}^{-3/2}-\bar{t}|^{4/3}d\Omega^2,\label{infinite1}\\
8\pi G\mu&=&\frac{-4}{3\bar{t}(s_1\bar{r}^{-3/2}-\bar{t})},\label{infinite2}\\
2Gm&=&\frac49 \bar{r}^3, \label{infinite3}
\end{eqnarray}
is obtained. We omit the bars on ${\bar t}$ and ${\bar r}$ hereafter for simplicity. This solution is an asymptotically flat FRW solution for $r \to \infty$ and belongs to the Lema{\^ i}tre-Tolman-Bondi family of solutions for the marginally bound case. We call this solution the infinite-kind KSS-LTB solution. It is noted that for $s_1 \ne 0$, this solution has no physical center, and $r=0$ corresponds to the $R=|s_1|^{2/3}$. For $s_1=0$, this solution gives the flat FRW solution. For this solution, the energy density is positive in the regions satisfying
\begin{eqnarray}
t<0, \quad s_1r^{-3/2}<t,~~\mbox{for} \quad s_1>0
\end{eqnarray}
and
\begin{eqnarray}
t<s_1r^{-3/2}, \quad 0<t,~~\mbox{for} \quad s_1<0.
\end{eqnarray}
The singular surfaces are $t=s_1r^{-3/2}$ and $t=0$, which corresponds to $R=|s_1|^{2/3}$. The former is a shell-focusing singularity, because the circumferential radius $R$ is zero there, while the latter is a shell-crossing singularity, because the quantity 
\begin{eqnarray}
\frac{\partial R}{\partial r}=
\left\{
\begin{array}{rl}
(-t)/(s_1r^{-3/2}-t)^{1/3}~~\mbox{for}~~t<s_1r^{-3/2} ,\\
t/(t-s_1r^{-3/2})^{1/3}~~\mbox{for}~~t>s_1r^{-3/2}
\end{array}\right.
\end{eqnarray}
is zero there. The fluid is collapsing in the region in which $t<s_1r^{-3/2}$ and expanding in the region in which $t>s_1r^{-3/2}$, because the quantity 
\begin{eqnarray}
\frac{\partial R}{\partial t}=-\mbox{sign}(s_1r^{-3/2}-t)\frac23r|s_1r^{-3/2}-t|^{-1/3}
\end{eqnarray}
is negative in the former case and positive in the latter case. The trapped region is defined by 
\begin{eqnarray}
s_1r^{-3/2}-(8/27)r^3<t<s_1r^{-3/2}
\end{eqnarray}
and
\begin{eqnarray}
s_1r^{-3/2}<t<s_1r^{-3/2}+(8/27)r^3.
\end{eqnarray}
The apparent horizons are 
\begin{eqnarray}
t=s_1r^{-3/2}+(8/27)r^3,~~(\mbox{AH1}) \label{AHinfinite1}
\end{eqnarray}
and
\begin{eqnarray}
t=s_1r^{-3/2}-(8/27)r^3.~~(\mbox{AH2}) \label{AHinfinite2}
\end{eqnarray}
Here we consider the trajectories of the apparent horizons. We consider only the solution with $\mu>0$ in which $R_r>0$ are satisfied. From Eqs.~(\ref{AHinfinite1}) and (\ref{AHinfinite2}), we obtain 
\begin{eqnarray}
\frac{dt}{dr}\big{|}_{\mbox{AH}}=
\left\{
\begin{array}{ll}
-(3s_1/2)r^{-5/2}+(8/9)r^2 \quad \mbox{for AH1},\\
-(3s_1/2)r^{-5/2}-(8/9)r^2 \quad \mbox{for AH2}.\\
\end{array}\right.
\end{eqnarray}
AH1 has its extremum at $r=r_7 \equiv (27s_0/16)^{2/9}$, while AH2 does not. For the null geodesics on the apparent horizon, the relation
\begin{eqnarray}
\frac{dt}{dr}\big{|}_{\pm}=
\left\{
\begin{array}{ll}
\pm|(3s_1/2)r^{-5/2}+(4/9) r^2|&\mbox{for AH1},\\
\pm|(3s_1/2)r^{-5/2}-(4/9) r^2|&\mbox{for AH2}\\
\end{array}\right.
\end{eqnarray}
are satisfied. In the region $r>0$, the trajectory of AH1 is spacelike (timelike) for
\begin{eqnarray}
r^{9/2}<(>)(27/4)s_1.
\end{eqnarray}
It is outgoing null for
\begin{eqnarray}
r^{9/2}=(27/4)s_1 \equiv r_8^{9/2}.
\end{eqnarray}
The trajectory of AH2 is timelike. Next, we consider $r=0$. The inequality $\mu<0$ is satisfied for $r \to 0$ along AH2, which is unphysical. Therefore we consider only AH1. For AH1, the relation
\begin{eqnarray}
\lim_{r \to 0}\left(\frac{dt}{dr}\big{|}_{\mbox{AH}}\big{/}\frac{dt}{dr}\big{|}_{\pm}\right)=\mp 1 \label{ratioinfinite}
\end{eqnarray}
is satisfied, so that AH1 is ingoing null at $r=0$.

\begin{figure}[htb]
\centerline{
\epsfxsize 11cm
\epsfbox{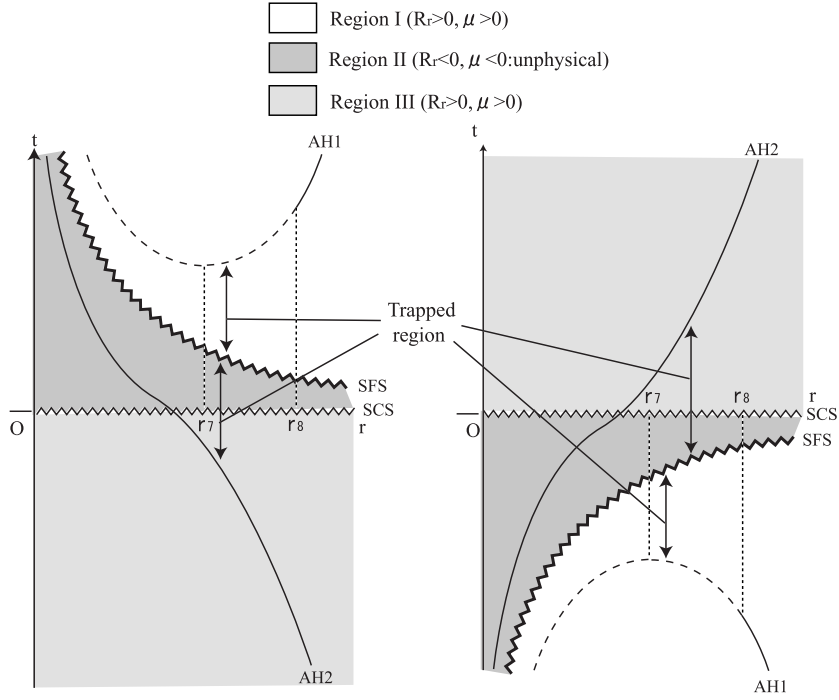}}
\vspace{0cm}
\caption{The figure on the left is for the infinite-kind KSS-LTB solution [Eqs~(\ref{infinite1})--(\ref{infinite3})], and that on the right is for its time-reversal solution. AH, SFS and SCS indicate an apparent horizon, a shell-focusing singularity and a shell-crossing singularity, respectively. Each diagram is divided by SFS and SCS into three regions, which we call regions I, II and III. Region II is unphysical, because $\mu<0$ there. A solid curve represents an apparent horizon whose trajectory is timelike, while a dashed curve represents an apparent horizon whose trajectory is spacelike. $r=r_7$ corresponds to the extremum of AH1. The trajectory of AH1 is outgoing (ingoing) null at $r=r_8$ in the left (right) figure. It is noted that $r \to +0$ and $r \to +\infty$ correspond to $R \to |s_1|^{2/3}$ (constant) and $R \to +\infty$, respectively.}
\label{infinite}
\end{figure}
A schematic figure for this solution is given in Fig.~\ref{infinite}, along with that for its time-reversal solution. The solution is unphysical in region II, in which the energy density is negative. Region I on the left in Fig.~\ref{infinite} represents a decaying white hole, while that in its time-reversal solution represents a growing black hole. Region III on the left in Fig.~\ref{infinite} represents the formation of a shell-crossing singularity at $t=0$ that corresponds to $R=|s_1|^{2/3}$ from a regular spacetime with no physical center. That in its time-reversal solution represents a shell-crossing singularity at $t=0$ that corresponds to $R=|s_1|^{2/3}$ and a regular spacetime with no physical center for $t>0$.

\subsection{$\xi^{\mu}$ parallel to $U^{\mu}$}
\label{sec:dustparallel}
In the parallel case for dust solutions of the second, zeroth and infinite kinds, $\Phi'=0$ is obtained from basic equations. Therefore we set $\exp(\Phi)=c_0$, where $c_0$ is a positive constant, in this subsection.

\subsubsection{Self-similarity of the second and zeroth kinds}
In these cases, the Einstein equations imply that the quantities $m$ and $\mu$ must be of the forms 
\begin{eqnarray}
2Gm&=&tM_1(r)+t^{3-2\alpha}M_2(r),\label{param}\\
8\pi G \mu&=&t^{-2}W_1(r)+t^{-2\alpha}W_2(r).\label{paramu}
\end{eqnarray}
The basic equations are the following:
\begin{eqnarray}
M_1&=&0,\label{para1}\\
(3-2\alpha)M_2&=&0,\label{para2}\\
M_1'&=&W_1S^2S',\label{para3}\\
M_2'&=&W_2S^2S',\label{para4}\\
M_1&=&S[1-S'^2],\label{para5}\\
M_2&=&c_0^{-2}S^3,\label{para6}\\
0&=&-W_1,\label{para7}\\
0&=&(2\alpha-3) W_2,\label{para8}\\
2S''S+S'^2&=&1-W_1S^2, \label{para00a}\\
3&=&c_0^2 W_2, \label{para00b}\\
S'^2&=&1, \label{para11a}\\
2\alpha-3&=&0, \label{para11b}
\end{eqnarray}
where the prime denotes the differentiation with respect to $r$. Equation~(\ref{para11b}) shows that $\alpha=3/2$ is the only case, and there are no solutions in the case of self-similarity of the zeroth kind. Then, $M_1=0$, $W_1=0$ and $W_2=3/c_0^2$ are obtained from Eqs. (\ref{para1}), (\ref{para7}) and (\ref{para00b}). Equation (\ref{para11a}) gives the evolution equation for $S$,
\begin{eqnarray}
S'^2=1,
\end{eqnarray}
which is integrated to yield
\begin{eqnarray}
S=\pm r+s_0,
\end{eqnarray}
where $s_0$ is an integration constant. The relation $M_2=S^3/c_0^2$ is obtained from Eq. (\ref{para6}). $S$ must be positive for the positive mass solution. The resulting solution is given by
\begin{eqnarray}
e^{2\Phi}&=&c_0^2,\\
R&=&t(s_0 \pm r),\\
8\pi G\mu&=&\frac{3}{c_0^2 t^3},\\
2Gm&=&c_0^{-2}(s_0 \pm r)^3.
\end{eqnarray}
This solution is the flat FRW solution. With the coordinate transformations $t \to \bar{t}=(2c_0/3)t^{3/2}$ and $r \to \bar{r}=(3/2c_0)^{2/3}(s_0 \pm r)$, the solution in the more usual form 
\begin{eqnarray}
ds^2&=&-d\bar{t}^2+\bar{t}^{4/3}(d\bar{r}^2+\bar{r}^2d\Omega^2),\\
8\pi G\mu&=&\frac{4}{3\bar{t}^2},\\
2Gm&=&\frac49 \bar{r}^3
\end{eqnarray}
is obtained.

\subsubsection{Self-similarity of the infinite kind}
In this case, the Einstein equations imply that the quantities $m$ and $\mu$ must be of the forms 
\begin{eqnarray}
2Gm&=&M(r),\\
8\pi G \mu&=&W(r),
\end{eqnarray}
and the Einstein equations and the equations of motion for the matter field (\ref{basic1})-(\ref{11}) are written
\begin{eqnarray}
M&=&S(1-S'^2),\label{parainf1}\\
M'&=&WS'S^2,\label{parainf2}\\
2SS''+S'^2&=&1-WS^2,\label{parainf00}\\
S'^2&=&1,\label{parainf11}
\end{eqnarray}
where the prime denotes the differentiation with respect to $r$. Because Eqs. (\ref{parainf00}) and (\ref{parainf11}) imply $W=0$, there are no solutions in this case.

\subsection{$\xi^{\mu}$ orthogonal to $U^{\mu}$}
\label{sec:dustorthogonal}
For the dust case, in the case of self-similarity of the second kind, Eqs.~(\ref{basic1}) and (\ref{orthfinite}) are contradictory, and in the case of self-similarity of the infinite kind, Eqs.~(\ref{basic1}) and (\ref{orthinfinite}) are contradictory, and therefore there are no solutions in these cases.

\subsubsection{Self-similarity of the zeroth kind}
In this case, the Einstein equations imply that the quantities $m$ and $\mu$ must be of the forms 
\begin{eqnarray}
2Gm&=&rM_1(t)+r^{3}M_2(t),\label{orthm}\\
8\pi G \mu&=&r^{-2}W_1(t)+W_2(t).\label{orthmu}
\end{eqnarray}
In this case, the basic equations are the following:
\begin{eqnarray}
M_1&=&W_1S^3,\label{orth1}\\
3M_2&=&W_2S^3,\label{orth2}\\
M_1'&=&0,\label{orth3}\\
M_2'&=&0,\label{orth4}\\
M_1&=&S(1-e^{-2\Psi}S^2),\label{orth5}\\
M_2&=&SS'^2 ,\label{orth6}\\
W_1'S&=&-W_1(\Psi'S+2S'),\label{orth9}\\
W_2'S&=&-W_2(\Psi'S+2S'),\label{orth10}\\
S'&=&S\Psi',\label{orth11}\\
-W_1S^2&=&e^{-2\Psi}S^2-1,\label{orth00a}\\
W_2S^2&=&S'^2+2\Psi'S'S,\label{orth00b}\\
0&=&1-S^2 e^{-2\Psi},\label{orth11a}\\
0&=&2S''S+S'^2,\label{orth11b}
\end{eqnarray}
where the prime denotes the differentiation with respect to $t$. We obtain $\exp(2\Psi)=S^2$ from Eq. (\ref{orth11a}), so that $M_1=0$ is concluded from Eq. (\ref{orth5}). The relation $M_1=0$ implies $W_1=0$, from Eq. (\ref{orth1}). Equation (\ref{orth11b}) gives the evolution equation for $S$,
\begin{eqnarray}
2y'+3y^2=0,
\end{eqnarray}
where $y \equiv S'/S$. This equation can be integrated to yield
\begin{eqnarray}
S=|s_1-s_0 t|^{2/3},
\end{eqnarray}
where $s_0$ and $s_1$ are integration constants. The equalities $M_2=SS'^2$ and $W_2=3y^2$ are obtained from Eqs. (\ref{orth6}) and (\ref{orth00b}), respectively. The resulting solution is 
\begin{eqnarray}
e^{2\Psi}&=&|s_1-s_0 t|^{4/3},\\
R&=&r|s_1-s_0 t|^{2/3},\\
8\pi G\mu&=&\frac{4s_0^2}{3(s_1-s_0 t)^2},\\
2Gm&=&\frac49 s_0^2r^3.
\end{eqnarray}
This solution is the flat FRW solution. With the coordinate transformations $t \to \bar{t}=s_1/s_0-t$ and $r \to \bar{r}=|s_0|^{2/3}r$, the solution in the more usual form 
\begin{eqnarray}
ds^2&=&-d\bar{t}^2+\bar{t}^{4/3}(d\bar{r}^2+\bar{r}^2d\Omega^2),\\
8\pi G\mu&=&\frac{4}{3\bar{t}^2},\\
2Gm&=&\frac49 \bar{r}^3
\end{eqnarray}
is obtained.

\section{Corrections and complements to I}
\label{sec:correction}
In this section, we give some corrections and complements to I.~\cite{mhio2002b} The first is only a typographical error: The right-hand side of Eq.~(4.44) in I should be replaced by $\alpha^{-2}t'^{(3-2\alpha)/\alpha}r'^3$. The others are more important: It is found that the solutions that we called the dynamical solutions (A) and (B) result in the singular static solution and the flat FRW solution, respectively. Although the analytic forms of the general solutions have not been obtained for the infinite kind in the tilted case with $p+\mu=0$ (Eqs.~(3.126)--(3.132) of I), they are found to be the Nariai solutions,~\cite{nariai} so that the solution, which is a special solution and we called the $\Lambda$-cylinder solution, is also the Nariai solution. We give a detailed demonstrations of them in the following. 

In the case $P_2=W_2=0$ and $\alpha=3/2$ considered in \S 3.1.2 of I, Eqs.~(3.15) and (3.55) there give 
\begin{eqnarray}
0&=&y[(1+K)(9K+2)y'+2(3K+1)(1+K)^2yy' \nonumber \\
&&+3(3K+1)(1+K)^2y^2+3(1+K)(9K+2)y+3(K^2+6K+1)]. \label{error0}
\end{eqnarray}
Eliminating $y'$ from Eq.~(\ref{error0}) by using Eq.~(3.51) of I, we obtain an algebraic equation for $y$:
\begin{eqnarray}
0=Ky^2[(1+K)y+1][6(1+K)(3K+1)y+(11+15K)]. \label{error1}
\end{eqnarray}
Then, because $K \ne 0$ has been assumed, $y$ must be constant. From Eq.~(3.51), the constant $y$ must be equal to $-(K-3)/[3(1-K)(1+K)]$ or $0$. Substituting $y=-(K-3)/[3(1-K)(1+K)]$ into Eq. (\ref{error1}), we obtain
\begin{eqnarray}
0=K^2(K-3)^2(3K+5)(3K-1),
\end{eqnarray}
and therefore $K=1/3$ is concluded. When $K=1/3$, which implies $y=-1$, Eq.~(3.55) of I gives the contradictory relation $\exp(2\Psi)=0$, and hence $y$ must be zero. The equality $y=0$ results in the singular static solution given by Eqs.~(3.30)--(3.35) of I. 

In the case $P_2=W_2=0$ and $K \ne -1$ considered in \S 3.3.2 of I, the relations 
\begin{eqnarray}
2\Phi'&=&-\frac{W_1'}{W_1}+\frac{S''}{S'}-\frac{S'}{S}, \label{error5}\\
2\Psi'&=&\left[\frac{2K}{1+K}(S'^2+SS'')+2(1+3K)S'S''\right]\nonumber\\
&&\big{/}\left[\frac{2K}{1+K}SS'+(1+3K)S'^2\right] \label{error6}
\end{eqnarray}
are obtained from Eqs.~(3.140) and (3.141) of I. Substituting Eqs.~(\ref{error5}) and (\ref{error6}) into Eq.~(3.115) of I and eliminating $y'$ by using Eq.~(3.139) of I, we obtain 
\begin{eqnarray}
K[3(1+3K)(1+K)y^2+(9K+5)y+2]=0.  \label{error7}
\end{eqnarray}
Because we have assumed $K \ne 0$, the solution of Eq.~(\ref{error7}) is $y=-2/[3(1+K)]$, which is consistent with Eq.~(3.139) of I for $K\ne 1$, while it is consistent with Eq.~(3.141) of I for $K=1$. This solution is found to be the flat FRW solution:
\begin{eqnarray}
e^{\Phi}&=&c_0,\\
e^{\Psi}&=&\frac{2}{3(1+K)}s_0 \xi^{-2/[3(1+K)]},\\
S&=&s_0 \xi^{-2/[3(1+K)]},\\
M_1&=&\frac{4}{9(1+K)^2} c_0^{-2}s_0^3 \xi^{-2/(1+K)},\\
M_2&=&0,\\
P_1&=&KW_1=\frac{4K}{3(1+K)^2} c_0^{-2},\\
P_2&=&KW_2=0,
\end{eqnarray}
where $c_0$ and $s_0$ are positive constants. With the coordinate transformations $t \to {\bar t}=c_0 t$ and $r \to {\bar r}=(s_0/c_0^{2/[3(1+K)]}) r^{-2/[3(1+K)]}$, a more usual form of the solution,
\begin{eqnarray}
ds^2&=&-d{\bar t}^2+{\bar t}^{4/[3(1+K)]}(d{\bar r}^2+{\bar r}^2d\Omega^2),\\
2Gm&=&\frac{4}{9(1+K)^2} \frac{{\bar r}^3}{{\bar t}^{2K/(1+K)}},\\
8\pi G p&=&8\pi G K\mu=\frac{4K}{3(1+K)^2}{\bar t}^{-2},
\end{eqnarray}
is obtained.

We show that a 4-dimensional spherically symmetric spacetime, which contains a perfect fluid with $p+\mu=0$ and has a constant circumferential radius, is described by the Nariai solution. In the following, Greek indices denote the 4-dimensional spacetime, uppercase Latin indices the 2-dimensional $(1+1)$ reduced spacetime, and lowercase Latin indices the 2-spheres. We write the general spherically symmetric spacetime as a manifold $M=M^2 \times S^2$ with the metric 
\begin{eqnarray}
g_{\mu\nu}=\mbox{diag}(g_{AB},R^2\gamma_{ab}),
\end{eqnarray}
where $g_{AB}$ is an arbitrary Lorentzian metric on $M^2$, $R$ is a scalar on $M^2$ and $\gamma_{ab}$ is the unit curvature metric on $S^2$. With the shorthand 
\begin{eqnarray}
v_A \equiv \frac{R_{|A}}{R},
\end{eqnarray}
where $|$ denotes the covariant derivative on $M^2$, the Einstein equations for a perfect fluid with $p+\mu=0$ in spherical symmetry, in the $2+2$ split, are
\begin{eqnarray}
-2(v_{A|B}+v_A v_B)+(2v_C^{~~|C}+3v_Cv^C-R^{-2})g_{AB}&=&-8\pi G \mu g_{AB}, \label{2+2a}\\
v_C^{~~|C}+v_C v^C-{\cal R}&=&-8\pi G \mu,\label{2+2b}
\end{eqnarray}
where ${\cal R} \equiv (1/2)R^A_{~~A}$ is the Gauss curvature of $g_{AB}$. The energy-momentum conservation equations give that $\mu$ is constant (see Eqs.~(\ref{b1}) and (\ref{b2})). When $R$ is constant, Eqs.~(\ref{2+2a}) and (\ref{2+2b}) give that $R^2=1/(8\pi G \mu)$ and ${\cal R}=8\pi G \mu$, i.e., $\mu$ must be positive and $M^2$ is a 2-dimensional positive constant curvature spacetime. Since a 2-dimensional positive constant curvature spacetime is the 2-dimensional de-Sitter spacetime ($dS^2$), the resulting 4-dimensional spacetime is the Nariai spacetime, which is $dS^2 \times S^2$ with a constant circumferential radius. Since the spherically symmetric kinematic self-similar perfect-fluid solutions of the infinite kind in the tilted case with $p+\mu=0$, which satisfies Eqs.~(3.127)--(3.132) of I, have a constant circumferential radius, they are found to be the Nariai solutions,~\cite{nariai} although the analytic forms of the general solutions in the self-similarity coordinates have not been obtained.

\section{Spherically symmetric kinematic self-similar perfect-fluid solutions of the first kind}
\label{sec:perfectfluid}
\subsection{Tilted case}
In this case, complete classification has already been carried out. (See Refs.~\citen{cc2000,gnu1998,ccgnu2000,ccgnu2001}.)

\subsection{$\xi^{\mu}$ parallel to $U^{\mu}$}
\label{sec:perfectfluidparallel}
In this case, the Einstein equations imply that the quantities $m, \mu$ and $p$ must be of the forms
\begin{eqnarray}
2Gm&=&tM(r),\\
8\pi G \mu&=&t^{-2}W(r),\\
8\pi G p&=&t^{-2}P(r).
\end{eqnarray}
The resulting equations for a perfect fluid, (\ref{basic1})--(\ref{11}), reduce to the following:
\begin{eqnarray}
P&=&-\frac13 W,\label{para11}\\
(P+W)\Phi'&=&-P',\label{para12}\\
M&=&-PS^3,\label{para13}\\
M'&=&WS^2S',\label{para14}\\
0&=&\Phi'S,\label{para15}\\
M&=&S(1+e^{-2\Phi}S^2-S'^2),\label{para16}\\
1-WS^2&=&-3S^2 e^{-2\Phi}+2SS''+S'^2, \label{para001}\\
-1-PS^2&=&S^2 e^{-2\Phi}-2\Phi'S'S-S'^2, \label{para111}
\end{eqnarray}
where the prime denotes the differentiation with respect to $r$. These equations result in 
\begin{eqnarray}
e^{\Phi}&=&c_0,\\
P&=&-\frac13 W=p_0,\\
M&=&-p_0S^3,\\
S'^2&=& 1+(c_0^{-2}+p_0)S^2,\\
S''&=&(c_0^{-2}+p_0)S,
\end{eqnarray}
where $c_0$ is a positive constant, and the constant $p_0$ must be negative for positive energy density. The resulting solution is the FRW solution:
\begin{eqnarray}
S=\left\{
\begin{array}{rl}
(1/a)\sin{(ar)},& \quad \mbox{for $c_0^{-2}+p_0<0$},\quad (\mbox{closed FRW~})\\
s_1 \pm r,&\quad  \mbox{for $c_0^{-2}+p_0=0$},\quad (\mbox{flat FRW})\\
(1/a)\sinh{(ar)},&\quad \mbox{for $c_0^{-2}+p_0>0$},\quad (\mbox{open FRW~})
\end{array}\right.
\end{eqnarray}
\begin{eqnarray}
2Gm&=&-p_0 S^3 t,\quad 8\pi G p=-8\pi G(\mu/3)=\frac{p_0}{t^2},
\end{eqnarray}
where $s_1$ is an integration constant, and $a \equiv \sqrt{-(c_0^{-2}+p_0)}$ for the closed FRW solution, while $a \equiv \sqrt{c_0^{-2}+p_0}$ for the open FRW solution. $S$ must be positive for a positive mass solution. For the closed and open FRW solutions, the coordinate transformations $t \to \bar{t}=c_0 t$ and $r \to \bar{r}=ar$ are used to write the solution in the more usual form
\begin{eqnarray}
ds^2&=&-d\bar{t}^2+\frac{\bar{t}^2}{c_0^2 a^2}(d\bar{r}^2+\bar{S}^2d\Omega^2),\\
2Gm&=&-\frac{p_0}{c_0 a^3} \bar{S}^3 \bar{t},\\
8\pi G p&=&-8\pi G(\mu/3)=\frac{p_0 c_0^2}{\bar{t}^2},\\
\bar{S}&=&\left\{
\begin{array}{rl}
\sin{\bar{r}}, \quad (\mbox{the closed FRW~}) \\
\sinh{\bar{r}}.\quad (\mbox{the open FRW~}) 
\end{array}\right.
\end{eqnarray}
For the flat FRW solution, the coordinate transformations $t \to \bar{t}=c_0 t$ and $r \to \bar{r}=(r+s_1)/c_0$ are used to write the solution in the more usual form
\begin{eqnarray}
ds^2&=&-d\bar{t}^2+\bar{t}^2(d\bar{r}^2+\bar{r}^2d\Omega^2),\\
2Gm&=&\bar{t}\bar{r}^3,\\
8\pi G p&=&-8\pi G(\mu/3)=-\frac{1}{\bar{t}^2}.
\end{eqnarray}

In the vacuum case (i.e., that in which $p_0=0$), the Minkowski solution in the Milne form,
\begin{eqnarray}
ds^2&=&-d\bar{t}^2+\bar{t}^2(d\bar{r}^2+\sinh^2\bar{r}d\Omega^2),
\end{eqnarray}
is obtained.

\subsection{$\xi^{\mu}$ orthogonal to $U^{\mu}$}
\label{sec:perfectfluidorthogonal}
\begin{figure}[htbp]
\centerline{(a)
\epsfxsize 9cm
\epsfbox{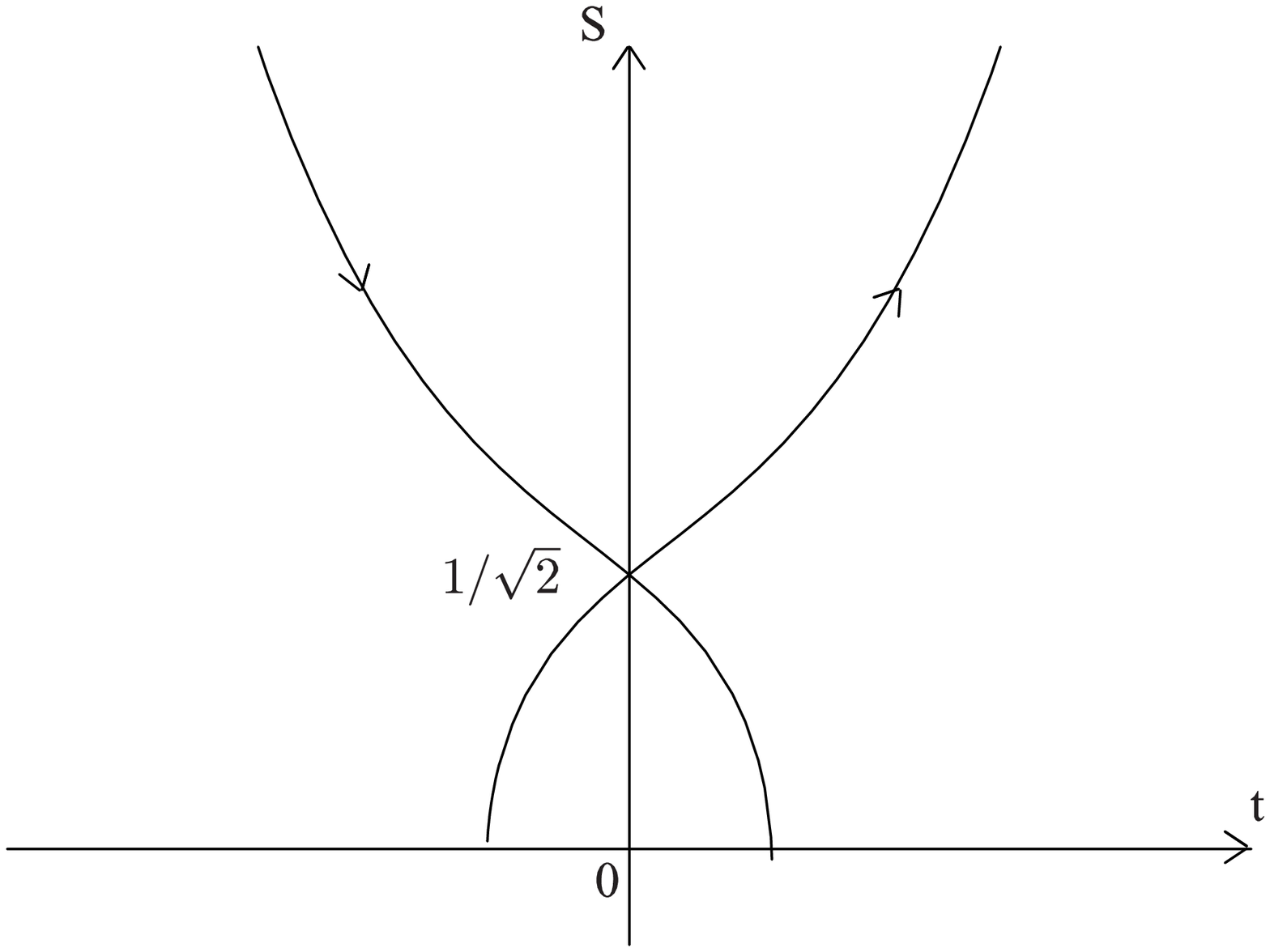}}
\centerline{(b)
\epsfxsize 9cm
\epsfbox{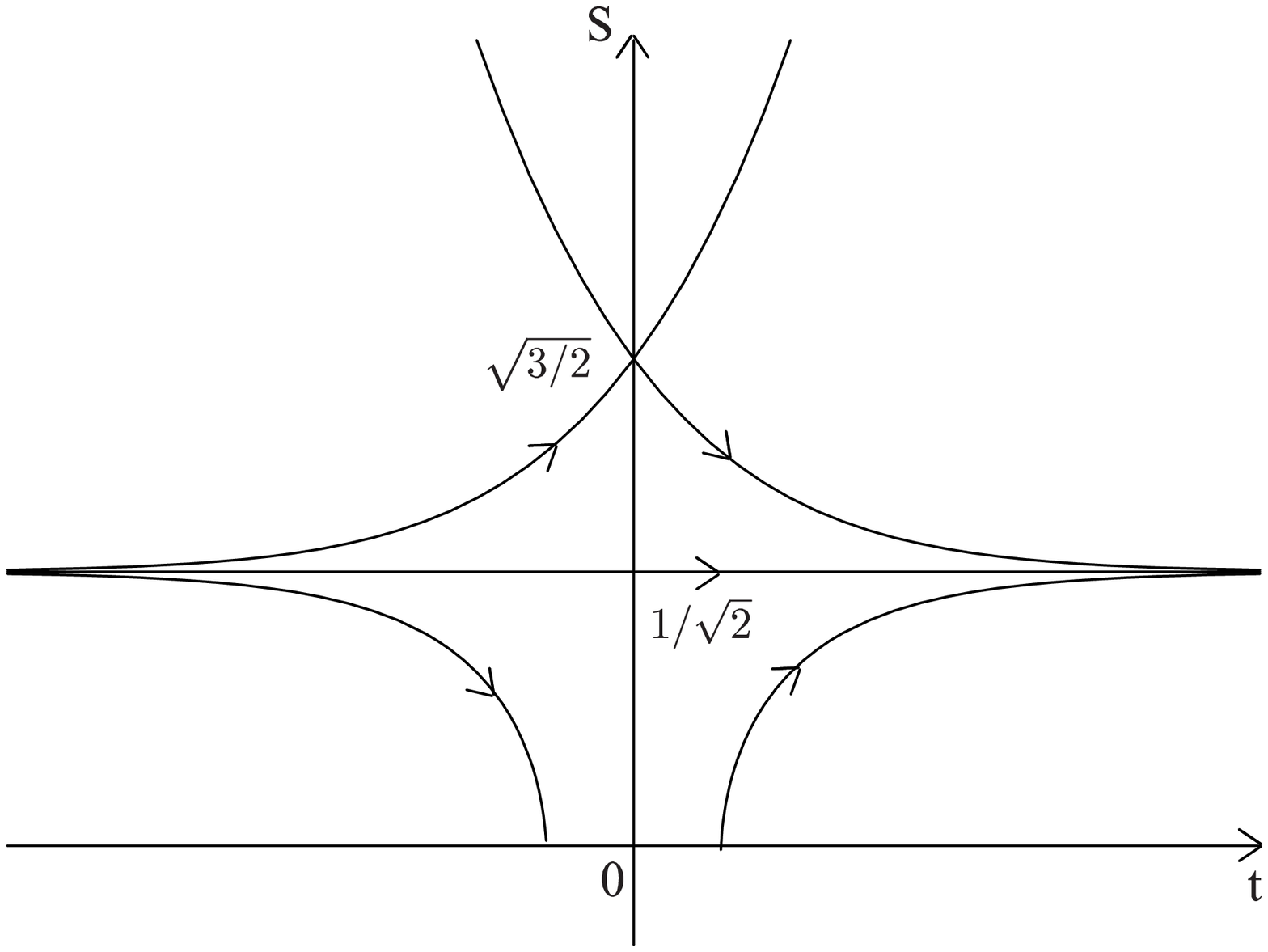}}
\centerline{(c)
\epsfxsize 9cm
\epsfbox{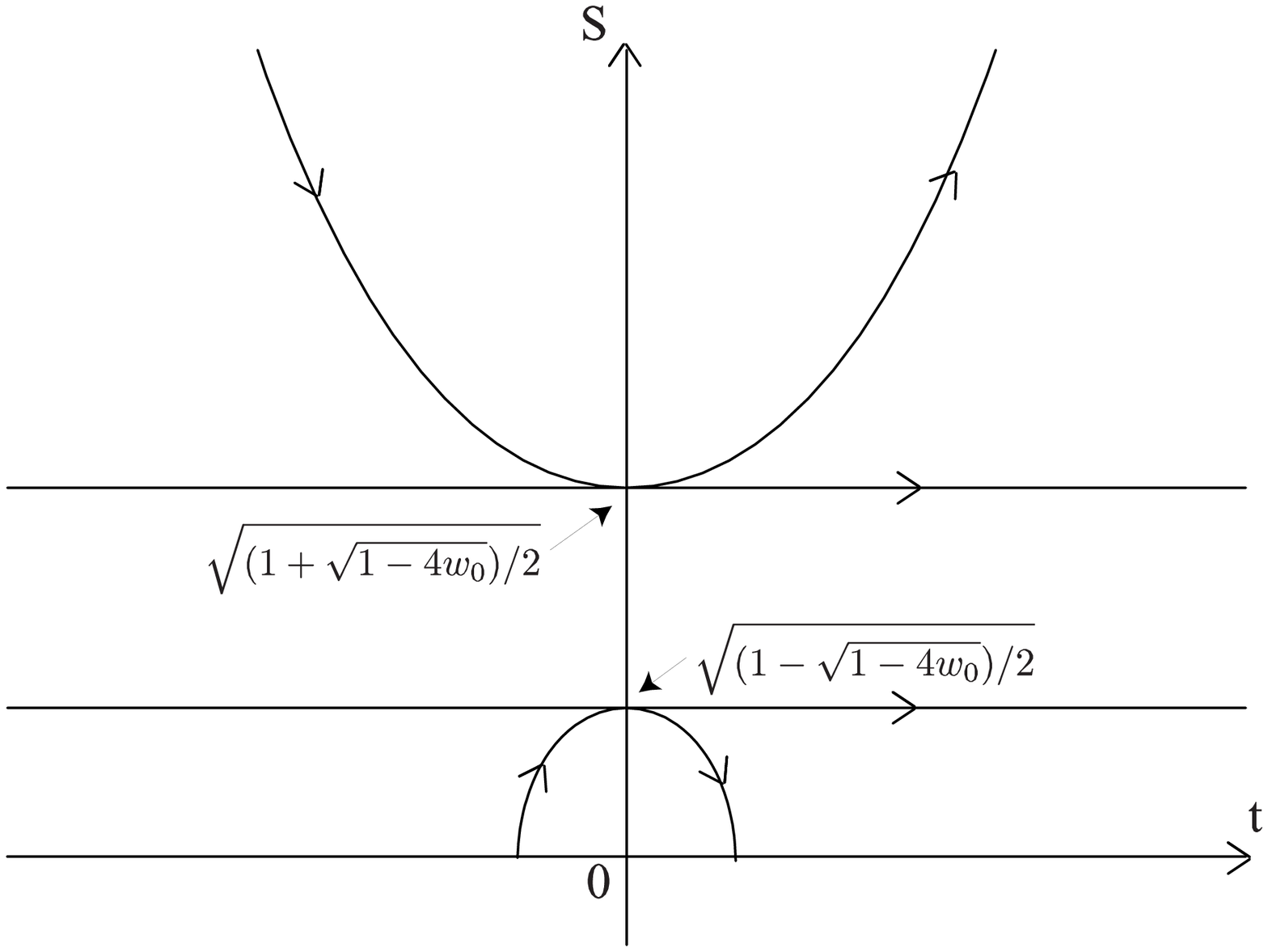}}
\vspace{0cm}
\caption{The function $S$ of the singular stiff-fluid solution for (a) $w_0 > 1/4$, (b) $w_0 = 1/4$ and (c) $0<w_0 < 1/4$. }
\label{stiffpic1}
\end{figure}

In this case, the Einstein equations imply that the quantities $m, \mu$ and $p$ must be of the forms
\begin{eqnarray}
2Gm&=&rM(t),\\
8\pi G \mu&=&r^{-2}W(t),\\
8\pi G p&=&r^{-2}P(t).
\end{eqnarray}
It is noted that in this case, the solution is always singular at the physical center. The resulting equations for a perfect fluid, (\ref{basic1})--(\ref{11}), reduce to the following:
\begin{eqnarray}
P&=&W,\label{orth11}\\
-W'S&=&(P+W)(\Psi'S+2S'),\label{orth12}\\
M&=&WS^3,\label{orth13}\\
M'&=&-PS^2S',\label{orth14}\\
0&=&\Psi'S,\label{orth15}\\
M&=&S(1+S'^2-e^{-2\Psi}S^2),\label{orth16}\\
1-WS^2&=&-S'^2+e^{-2\Psi}S^2,\label{orth100}\\
-1-PS^2&=&2S''S+S'^2-3e^{-2\Psi}S^2,\label{orth111}
\end{eqnarray}
where the prime denotes the differentiation with respect to $t$. These equations result in
\begin{eqnarray}
e^{\Psi}&=&c_1,\label{stiff1}\\
M&=&WS^3,\\
P&=&W=w_0 S^{-4},\\
S'&=&\pm \frac{\sqrt{c_1^{-2}S^4-S^2+w_0}}{S},
\end{eqnarray}
where $c_1$ and $w_0$ are positive constants. A positive value of $w_0$ is required for a positive energy density. $c_1$ can be set to unity with the transformations $c_1 r\to r$, $t/c_1 \to t$, $S/c_1 \to S$ and $w_0/c_1^2 \to w_0$. In this case, a general solution parametrized by $w_0$ can be obtained. The function $S$ is given by  
\begin{eqnarray}
S^2=1/2 \pm \sqrt{w_0-1/4}\sinh(2t+t_0)
\end{eqnarray}
for $w_0>1/4$, 
\begin{eqnarray}
S^2=1/2 \pm \sqrt{1/4-w_0}\cosh(2t+t_0),~\mbox{or}~~(1\pm\sqrt{1-4 w_0})/2
\end{eqnarray}
for $0<w_0<1/4$, and 
\begin{eqnarray}
S^2=1/2 +e^{\pm2t+t_0},~~1/2 -e^{\pm2t+t_0},~~\mbox{or}~~1/2 \label{stiff2}
\end{eqnarray}
for $w_0=1/4$, where $t_0$ is a constant and can be set to zero with the transformation $\pm2t+t_0 \to \pm2t$. We call this solution the ``singular stiff-fluid solution.'' The inequality $w_0>0$ must be satisfied for a positive energy density. Also, the sign of $S$ must be positive for a positive mass solution. The function $S$ in various cases is plotted in Fig.~\ref{stiffpic1}.
For $0<w_0 \le 1/4$, there exist static solutions with singular centers. Next, we discuss the dynamical solutions. For $w_0>1/4$, the dynamical solution represents a big-bang cosmological model or gravitational collapse with a singular center. For $0<w_0<1/4$, this solution represents a bouncing cosmological solution with a singular center or a big-bang cosmological model with a singular center that ends in a big-crunch. For $w_0=1/4$, the solution $S^2=1/2 +\exp(\pm2t)$ represents an expanding or contracting universe with a singular center, while the solution $S^2=1/2 -\exp(\pm2t)$ represents a big-bang cosmological model or gravitational collapse with a singular center. The trapping condition $2Gm<R$ corresponds to $w_0>S^2$.

In the vacuum case, these equations result in 
\begin{eqnarray}
e^{\Psi}&=&c_1,\\
S'^2&=& -1+c_1^{-2}S^2,\\
S''&=&c_1^{-2}S,
\end{eqnarray}
where $c_1$ is a constant. The solution is $S=c_1 \cosh (c_1^{-1} t)$, which is the Minkowski spacetime with the metric
\begin{eqnarray}
ds^2=-r^2dt^2+c_1^2dr^2+c_1^2 r^2\cosh^2(c_1^{-1} t)d\Omega^2.
\end{eqnarray}
This metric can be obtained with the coordinate transformations $(t,r) \to (\tau, x)$, with $\tau=c_1 r \sinh(c_1^{-1}t)$ and $x=c_1 r \cosh(c_1^{-1}t)$ from $ds^2=-d\tau^2+dx^2+x^2d\Omega^2$.

\section{Summary and discussion}
\label{sec:summary}

\begin{table}[htbp]
	\caption{Kinematic self-similar solutions for the dust case. Here, $\parallel$ and $\perp$ denote the parallel and orthogonal cases, respectively.}
	\label{tablesol}
	\begin{center}
		\begin{tabular}{|l|l|c|cccc}
		\hline\hline
		Self-similarity & Solution & Equation number \\ 
		\noalign{\hrule height 0.8pt}
		1st (tilted) & see Ref.~\citen{carr2000} & \\ \hline
		1st ($\parallel$) & none & \\ \hline 
		1st ($\perp$) & none &\\ \hline 
		2nd (tilted) & 2nd-kind KSS-LTB (see also Refs.~\citen{ch1989,bc1998,blvw2002})& (\ref{solsecond1})--(\ref{solsecond3})\\ \hline
		2nd ($\parallel$) & flat FRW ($\alpha=3/2$) &\\ \hline 
		2nd ($\perp$) & none & \\ \hline 
		zeroth (tilted) & zeroth-kind KSS-LTB (see also Refs.~\citen{bc1998,blvw2002}) & (\ref{solzero1})--(\ref{solzero3})\\ \hline 
		zeroth ($\parallel$) & none &\\ \hline 
		zeroth ($\perp$) & flat FRW  &\\ \hline 
		infinite (tilted) & infinite-kind KSS-LTB & (\ref{infinite1})--(\ref{infinite3})\\ \hline 
		infinite ($\parallel$) & none & \\ \hline
		infinite ($\perp$) & none &\\ 
		\noalign{\hrule height 0.8pt}
		\end{tabular}
	\end{center}
\end{table}

We have classified kinematic self-similar spherically symmetric dust solutions of the second, zeroth and infinite kinds and perfect-fluid solutions, together with vacuum solutions of the first kind, in which the kinematic self-similarity vector is either parallel or orthogonal to the fluid flow. Except for the cases in which there are no solutions, the governing equations can be integrated to obtain exact solutions. These kinematic self-similar dust solutions belong to the Lema{\^ i}tre-Tolman-Bondi family of solutions only for the marginally bound case, while, in contrast, there exist some first-kind dust solutions that belong to this family for the non-marginally bound case. We give a proof of this in Appendix~\ref{intcond}. We also presented corrections and complements to I.~\cite{mhio2002b} It is found that the solutions which in I we called the dynamical solutions (A) and (B) result in the singular static solution and the flat FRW solution, respectively. It is found that the spherically symmetric kinematic self-similar perfect-fluid solutions of the infinite kind in the tilted case with $p+\mu=0$, which satisfies Eqs.~(3.127)--(3.132) of I and includes the special solution which in I we called the $\Lambda$-cylinder solution, are the Nariai solutions.~\cite{nariai} The results of the present work, together with the corrected results of I are listed in Tables \ref{tablesol}--\ref{tablesol4}. 

We first summarize the results of this work for the first-kind cases. In the parallel case, the flat, open and closed FRW solutions with $p=-\mu/3$ are the only solutions. In the orthogonal case, the equation of state must be $p=\mu$, and a new exact solution, which we call the singular stiff-fluid solution, is the only solution. This solution is singular at the physical center. 

Next, we summarize the results for the dust case. There are no dust solutions in the parallel case of the second kind (except for $\alpha=3/2$), zeroth kind and infinite kind and in the orthogonal case of the second kind and infinite kind. The flat FRW solution is the second-kind solution in the parallel case for $\alpha=3/2$ and also the zeroth-kind solution in the orthogonal case. In the tilted case, general solutions, which represent a collapsing or an expanding dust fluid, can be obtained in closed form in the cases of the second kind (second-kind KSS-LTB solution), zeroth kind (zeroth-kind KSS-LTB solution) and infinite kind (infinite-kind KSS-LTB solution). The second-kind KSS-LTB solution with $\alpha=3/2$ is the flat FRW solution. The second-kind KSS-LTB solution with $\alpha \ne 3/2$, the zeroth-kind KSS-LTB solution and the infinite-kind KSS-LTB solution are characterized by a single parameter, and the second-kind KSS-LTB solution for $\alpha \ne 3/2$ and the infinite-kind KSS-LTB solution include the flat FRW solution as a special case. In all three solutions, if they are not the flat FRW solutions, there exist a shell-crossing singularity and a shell-focusing singularity. There exists an unphysical region, in which the energy density is negative. We considered only physical regions. In the collapsing regions, the second-kind KSS-LTB solutions for $0<\alpha<1$ and $1<\alpha<3/2$ represent the formation of a shell-focusing singularity from a regular spacetime. For $1<\alpha<3/2$, this singularity is covered; i.e., this solution represents the formation and growth of a black hole. It is important to clarify whether the singularity is naked or covered for $0<\alpha<1$. The second-kind KSS-LTB solution for $\alpha<0$ and $3/2<\alpha$, the zeroth-kind KSS-LTB solution and the infinite-kind KSS-LTB solution represent a growing black hole in the collapsing region, although they do not represent the formation of a black hole. These solutions could be important in modeling the growth of a black hole. The second-kind KSS-LTB solution for $0<\alpha<1$ and $1<\alpha<3/2$ in the expanding region could be important in modeling the evolution of a cosmic void. The infinite-kind KSS-LTB solution with $s_1 \ne 0$ does not have a physical center. This solution could be attached to an inner region that is represented by another solution. This solution could also be important in modeling dust collapse or the evolution of a cosmic void. In the high-density region, a relativistic polytrope gas might be approximated by a dust, so that the central region in the generic collapse of a gas could converge to one of these kinematic self-similar solutions. 

Adding the results of the present work to those of Refs.~\citen{carr2000,cc2000,mhio2002a,mhio2002b}, the classification of spherically symmetric kinematic self-similar solutions with the vacuum, dust or perfect fluid obeying a relativistic polytropic equation of state or that of the form $p=K\mu$, has been completed.

\begin{table}[htbp]
		\caption{Kinematic self-similar solutions for an equation of state of the form $p=K \mu^{\gamma}$ [EOS(I)], where $K$ and $\gamma$ are constants. It is assumed that $K \ne 0$ and $\gamma \ne 0,1$. Here, $\parallel$ and $\perp$ denote the parallel and orthogonal cases, respectively. (See I~\cite{mhio2002b} for the values of the parameters.)}
	\label{tablesol1}
		\begin{center}
		\begin{tabular}{|l|l|ccccc}
		\hline\hline
		Self-similarity & Solution \\ 
		\noalign{\hrule height 0.8pt}
		1st (tilted) & none \\  \hline 
		1st ($\parallel$) & none   \\  \hline 
		1st ($\perp$) & none  \\  \hline 
		2nd (tilted) & none  \\  \hline 
		2nd ($\parallel$) & none   \\  \hline 
		2nd ($\perp$) & none  \\  \hline 
		zeroth (tilted) & none  \\  \hline 
		zeroth ($\parallel$) & none  \\ \hline 
		zeroth ($\perp$) & flat FRW  \\ \hline 
		infinite (tilted) & none  \\ \hline 
		infinite ($\parallel$) & All static solutions with the EOS (I)  \\ \hline
		infinite ($\perp$) & none  \\ 
		\noalign{\hrule height 0.8pt}
		\end{tabular}
	\end{center}
\end{table}
\begin{table}[htbp]
		\caption{Kinematic self-similar solutions for an equation of state of the form that $p=K n^{\gamma}$ and $\mu=m_b n+p/(\gamma-1)$ [EOS(II)], where the constant $m_b$ and $n(t,r)$ represent the mean baryon mass and the baryon number density, respectively, and $K$ and $\gamma$ are constants. It is assumed that $K \ne 0$ and $\gamma \ne 0,1$. Here, $\parallel$ and $\perp$ denote the parallel and orthogonal cases, respectively. (See I~\cite{mhio2002b} for the values of the parameters.)}
	\label{tablesol2}
		\begin{center}
		\begin{tabular}{|l|l|ccccc}
		\hline\hline
		Self-similarity & Solution  \\ 
		\noalign{\hrule height 0.8pt}
		1st (tilted) & none \\  \hline 
		1st ($\parallel$) & none  \\  \hline 
		1st ($\perp$) & none \\  \hline 
		2nd (tilted) & none  \\  \hline 
		2nd ($\parallel$) & closed FRW  \\ \hline
		2nd ($\perp$) & none \\ \hline 
		zeroth (tilted) & none \\ \hline 
		zeroth ($\parallel$) & none \\ \hline 
		zeroth ($\perp$) & flat FRW  \\ \hline 
		infinite (tilted) & none \\ \hline 
		infinite ($\parallel$) & All static solutions with the EOS (II)   \\ \hline
		infinite ($\perp$) & none \\ 
		\noalign{\hrule height 0.8pt}
		\end{tabular}
	\end{center}
\end{table}
\begin{table}[htbp]
		\caption{Kinematic self-similar solutions for $p=K\mu$ [EOS(III)], where $K$ is a constant and $-1 \le K \le 1~(K \ne 0)$. Here, $\parallel$ and $\perp$ denote the parallel and orthogonal cases, respectively. (See the main text and I~\cite{mhio2002b} for the values of the parameters $\alpha$ and $K$.) It is noted that the kinematic self-similar solution of the infinite kind with $K=-1$ for a tilted case is the Nariai solution~\cite{nariai}, although the analytic forms of the solutions in the self-similarity coordinates have not been obtained. }
	\label{tablesol3}
	\begin{center}
		\begin{tabular}{|l|l|c|cccc}
		\hline \hline
		Self-similarity & Solution & Equation number \\ 
		\noalign{\hrule height 0.8pt}
		1st (tilted) & see Refs.~\citen{cc2000,op1990} & \\  \hline 
		1st ($\parallel$) & FRW ($K=-1/3$) & \\  \hline 
		1st ($\perp$) & Singular stiff-fluid ($K=1$)& (\ref{stiff1})--(\ref{stiff2}) \\  \hline 
		2nd (tilted) & flat FRW & \\
		             & Singular static & (3.41)--(3.46) in I \\ \hline
		2nd ($\parallel$) & flat FRW & \\ \hline
		2nd ($\perp$) & Singular static & (5.28)--(5.31) in I \\ \hline 
		zeroth (tilted) & de-Sitter ($K=-1$) & \\ 
		             & Singular static & (3.96)--(3.101) in I \\ \hline
		zeroth ($\parallel$) & de-Sitter ($K=-1$) & \\ \hline 
		zeroth ($\perp$) & flat FRW & \\ 
		             & de-Sitter ($K=-1$) & \\ \hline
		infinite (tilted) & flat FRW & \\
		             & Nariai ($K=-1$) & (3.127)--(3.132) in I \\ \hline
		infinite ($\parallel$) & All static solutions  with the EOS (III)&  \\ \hline
		infinite ($\perp$) & none & \\ 
		\noalign{\hrule height 0.8pt}
 
		\end{tabular}
	\end{center}
\end{table}
\begin{table}[htbp]
	\caption{Kinematic self-similar solutions for the vacuum case. Here, $\parallel$ and $\perp$ denote the parallel and orthogonal cases, respectively.}
	\label{tablesol4}
	\begin{center}
		\begin{tabular}{|l|l|ccccc}
		\hline \hline
		Self-similarity & Solution  \\ 
		\noalign{\hrule height 0.8pt}
		1st ($\parallel$) & Minkowski \\ \hline 
		1st ($\perp$) & Minkowski \\ \hline 
		2nd (tilted) & Minkowski (all $\alpha$)  \\
		             & Schwarzschild ($\alpha=3/2$) \\ \hline
		2nd ($\parallel$) & none \\ \hline 
		2nd ($\perp$) & none \\ \hline 
		zeroth (tilted) & Minkowski  \\ \hline 
		zeroth ($\parallel$) & none \\ \hline 
		zeroth ($\perp$) & Minkowski  \\ \hline 
		infinite (tilted) & none \\ \hline 
		infinite ($\parallel$) & Minkowski  \\
		             & Schwarzschild  \\ \hline
		infinite ($\perp$) & none \\ 
		\noalign{\hrule height 0.8pt}
		\end{tabular}
	\end{center}
\end{table}

\acknowledgements

The authors would like to thank H.~Kodama for critical comments on the Nariai solution, and K.~Maeda for continuous encouragement. This work was partly supported by Grants-in-Aid for Scientific Research (Nos.~05540 and 11217)
from the Japanese Ministry of
Education, Culture, Sports, Science and Technology.

\appendix

\section{Flatness of 3-Surfaces for Kinematic Self-Similar Dust Solutions}
\label{intcond}

In this appendix, we give the integrability conditions for kinematic self-similarity of the Einstein equations and a proof of the flatness of 3-surfaces for kinematic self-similar dust solutions. The covariant derivative of $u_{\mu}$ can be decomposed into its irreducible parts as
\begin{eqnarray}
u_{\mu;\nu} = \sigma_{\mu\nu} + \frac{1}{3} \theta h_{\mu\nu} + \omega_{\mu\nu} - \dot{u}_{\mu} u_{\nu},\label{b0}
\end{eqnarray}
where $\theta \equiv g^{\mu\nu} \theta_{\mu\nu}$ (with $\theta_{\mu\nu} \equiv h_{(\mu}^{~~\kappa}h_{\nu)}^{~~\lambda} u_{\kappa;\lambda}$) is the expansion scalar, $\sigma_{\mu\nu} \equiv \theta_{\mu\nu}-(1/3)\theta h_{\mu\nu}$ is the shear tensor (from which we obtain $\sigma^2 \equiv (1/2)\sigma_{\mu\nu}\sigma^{\mu\nu}$), $\omega_{\mu\nu} \equiv h_{[\mu}^{~~\kappa}h_{\nu]}^{~~\lambda}u_{\kappa;\lambda}$ is the vorticity tensor (from which we obtain $\omega^2 \equiv (1/2) \omega_{\mu\nu} \omega^{\mu\nu}$), and $\dot{u}_\mu \equiv u_{\mu;\nu} u^{\nu}$ is the acceleration vector.~\cite{ellis} The conservation laws become
\begin{eqnarray}
\dot{\mu} + (\mu + p) \theta &=&0, \label{b1}\\
(\mu +p) \dot{u}_{\mu} &=& -p_{,\nu}h^{\nu}_{~~\mu},\label{b2}
\end{eqnarray}
where the dot represents operation with ${}_{;\nu}u^{\nu}$, while the Ricci identities with the Einstein equations become
\begin{eqnarray}
&&\dot{\theta}+\frac{1}{3}\theta^2-\dot{u}^{\mu}_{~~;\mu}+2(\sigma^2-\omega^2)+\frac{8\pi G}{2}(\mu + 3p)=0,\label{b3}\\
&&h^{\mu}_{~~\nu}\left(\dot{\omega}^{\nu}+\frac{2}{3}\theta \omega^{\nu}\right)=\sigma^{\mu}_{~~\nu}\omega^{\nu}+\frac{1}{2}\eta^{\mu\nu\kappa\lambda} u_{\nu} \dot{u}_{\kappa;\lambda}, \label{b4}\\
&&h^{\mu}_{~~\nu} \left(\omega^{\nu\kappa}_{~~~~;\kappa}-\sigma^{\nu\kappa}_{~~~~;\kappa}+\frac{2}{3} \theta^{,\nu}\right)+(\omega^{\mu}_{~~\nu}+\sigma^{\mu}_{~~\nu}) \dot{u}^{\nu}=0,\label{b5}\\
&&\omega^{\mu}_{~~;\mu}=2\omega^{\mu} \dot{u}_{\mu},\label{b6}\\
&&h_{\mu}^{~~\kappa}h_{\nu}^{~~\lambda} \dot{\sigma}_{\kappa\lambda}-h_{\mu}^{~~\kappa}h_{\nu}^{~~\lambda}\dot{u}_{(\kappa;\lambda)}-\dot{u}_{\mu} \dot{u}_{\nu}+\omega_{\mu} \omega_{\nu}+\sigma_{\mu\kappa} \sigma^{\kappa}_{~~\nu}\nonumber \\
&&~~~~~~~~~~~~~~~~+\frac{2}{3} \theta \sigma_{\mu\nu}+h_{\mu\nu} \left(-\frac{1}{3}\omega^2-\frac{2}{3}\sigma^2+\frac{1}{3}\dot{u}^{\kappa}_{~~;\kappa} \right)+E_{\mu\nu}=0,\label{b7}\\
&&H_{\mu\nu}=2 \dot{u}_{(\mu} \omega_ {\nu)}-h_{\mu}^{~~\lambda}h_{\nu}^{~~\zeta} (\omega_{(\lambda}^{~~\nu;\rho}+\sigma_{(\lambda}^{~~\nu;\rho}) \eta_{\zeta)\kappa\nu\rho} u^{\kappa},\label{b8}
\end{eqnarray}
with
\begin{eqnarray}
E_{\mu\nu} &\equiv&C_{\mu\kappa\nu\lambda}u^\kappa u^\lambda,\\
H_{\mu\nu} &\equiv&\frac12 \eta_{\mu\kappa}^{~~~\rho\sigma}C_{\rho\sigma\nu\lambda}u^\kappa u^\lambda,
\end{eqnarray}
where $C_{\mu\nu\kappa\lambda}$ and $\eta_{\mu\nu\kappa\lambda}$ are the Weyl tensor and the totally-skew pseudotensor, respectively. In the case of zero vorticity ($\omega_{\mu\nu} =0$), the Gauss-Codacci equations become
\begin{eqnarray}
{}^3R_{\mu\nu}&=& (-\theta \sigma_{\kappa\lambda}-\dot{\sigma}_{\kappa\lambda}+\dot{u}_{(\kappa;\lambda)}) h_{\mu}^{~~\kappa}h_{\nu}^{~~\lambda}  \nonumber \\
&&+\dot{u}_{\mu}\dot{u}_{\nu}+\frac{1}{3} h_{\mu\nu} \left(-\frac{2}{3} \theta^2 + 2\sigma^2 + 16\pi G \mu - \dot{u}^{\rho}_{~~;\rho}\right),\label{gausseq} \\
{}^3R &=& -\frac{2}{3} \theta^2 + 2 \sigma^2 + 16\pi G \mu,\label{codaccieq}
\end{eqnarray}
where ${}^3R_{\mu\nu}$ is the Ricci tensor of the $3$-spaces orthogonal to $u^{\mu}$, and ${}^3R$ is the corresponding Ricci scalar.

Now we derive the integrability conditions for kinematic self-similarity on the Einstein equations, i.e., 
\begin{eqnarray}
{\cal L}_{\bf{\xi}} R_{\mu\nu} = 8\pi G {\cal L}_{\bf{\xi}}\left(T_{\mu\nu} -\frac12 g_{\mu\nu}T\right), \label{intcondition}
\end{eqnarray}
where $R_{\mu\nu}$ is the Ricci tensor and $\xi^{\mu}$ is a kinematic self-similarity vector that is defined by Eqs.~(\ref{kss}) and (\ref{gss}). From Eqs. (\ref{kss}) and (\ref{gss}), we obtain
\begin{eqnarray}
{\cal L}_{\bf{\xi}} g_{\mu\nu} = 2\delta g_{\mu\nu} + 2(\delta-\alpha) u_{\mu} u_{\nu}. 
\end{eqnarray}
Hence, we obtain
\begin{eqnarray}
{\cal L}_{\bf{\xi}} R_{\mu\nu}&=&({\cal L}_{\bf{\xi}} \Gamma^\kappa_{\mu\nu})_{;\kappa} - ({\cal L}_{\bf{\xi}} \Gamma^\kappa_{\mu\kappa})_{;\nu} \nonumber  \\
&=&(\delta-\alpha)\left[2 \dot{\sigma}_{\mu\nu} + 2 \theta \sigma_{\mu\nu} + 2 \sigma_{\mu\kappa} \omega^\kappa_{~~\nu} +2 \sigma_{\nu\kappa} \omega^\kappa_{~~\mu}+4\omega_{\mu\kappa} \omega^\kappa_{~~\nu}\right. \nonumber \\
&&+\frac{2}{3} \theta g_{\mu\nu} (\dot{\theta}+\theta)+u_\mu (2 \omega_{\kappa\nu}^{~~~;\kappa}-2 \sigma_{\nu\kappa} \dot{u}^\kappa-2\omega_{\nu\kappa} \dot{u}^\kappa) \nonumber \\
&& +u_\nu (2 \omega_{\kappa\mu}^{~~~;\kappa}-2 \sigma_{\mu\kappa} \dot{u}^\kappa-2\omega_{\mu\kappa} \dot{u}^\kappa) \nonumber \\
&& \left.+u_\mu u_\nu \left( \frac{2}{3} \theta (\dot{\theta} + \theta)-2\dot{u}_\kappa^{~~;\kappa} \right)\right], 
\end{eqnarray}
where
\begin{eqnarray}
{\cal L}_{\bf{\xi}} \Gamma^\mu_{~~\nu\lambda} \equiv \frac{1}{2} g^{\mu\kappa} [({\cal L}_{\bf{\xi}} g_{\kappa\nu})_{;\lambda} + ({\cal L}_{\bf{\xi}} g_{\kappa\lambda})_{;\nu} - ({\cal L}_{\bf{\xi}} g_{\nu\lambda})_{;\kappa}]. 
\end{eqnarray}
Decomposing this equation in various ways, we obtain
\begin{eqnarray}
({\cal L}_{\bf{\xi}} R_{\mu\nu})u^\mu u^\nu &=& (\delta-\alpha)(-8\omega^2 - 2 \dot{u}_\kappa^{~~;\kappa}), \label{lie1}\\
({\cal L}_{\bf{\xi}} R_{\mu\nu})h^{\mu\nu} &=& 2(\delta-\alpha)(\dot{\theta} + \theta^2 - 4 \omega^2), \\
({\cal L}_{\bf{\xi}} R_{\mu\nu})u^\mu h^\nu_{~~\lambda}&=& (\delta-\alpha)(2 \omega_{\lambda\mu} \dot{u}^\mu+2\omega_{\kappa\lambda}^{~~~;\kappa}\nonumber \\
&&-4\omega^2 u_\lambda), \\
({\cal L}_{\bf{\xi}} R_{\mu\nu}) \left(h^\mu_{~~\lambda} h^\nu_{~~\rho} - \frac{1}{3} h_{\lambda\rho} h^{\mu\nu}\right)  &=& 2(\delta-\alpha) \left(\dot{\sigma}_{\lambda\rho}-u_\rho \sigma_{\lambda\nu} \dot{u}^\nu - u_\lambda \sigma_{\rho\mu} \dot{u}^\mu \right.\nonumber \\
&&+ \theta \sigma_{\lambda\rho} + \sigma_{\lambda\kappa} \omega^\kappa_{~~\rho}+\sigma_{\rho\kappa} \omega^\kappa_{~~\lambda} \nonumber \\
&&\left.+ 2 \omega_\lambda^{~~\kappa} \omega_{\kappa\rho} + \frac{4}{3} h_{\lambda\rho}\omega^2 \right). \label{lie2}
\end{eqnarray}
In the perfect-fluid case, the relation
\begin{eqnarray}
T_{\mu\nu} -\frac12 g_{\mu\nu}T = \frac{1}{2} (\mu + 3p) u_\mu u_\nu + \frac{1}{2} (\mu -p) h_{\mu\nu} 
\end{eqnarray}
is satisfied. Then, we obtain
\begin{eqnarray}
{\cal L}_{\bf{\xi}} \left(T_{\mu\nu} -\frac12 g_{\mu\nu}T\right) &=& \frac{1}{2}\{{\cal L}_{\bf{\xi}}\mu+ 2 \alpha \mu + 3({\cal L}_{\bf{\xi}}p+ 2 \alpha p)\} u_\mu u_\nu \nonumber \\
&&+ \frac{1}{2} \{{\cal L}_{\bf{\xi}}\mu+2\delta \mu - {\cal L}_{\bf{\xi}}p-2\delta p\} h_{\mu\nu}. 
\end{eqnarray}
Decomposing this equation, we obtain
\begin{eqnarray}
\left[{\cal L}_{\bf{\xi}} \left(T_{\mu\nu} -\frac12 g_{\mu\nu}T\right)\right]u^\mu u^\nu  &=& \frac12({\cal L}_{\bf{\xi}}\mu + 2 \alpha\mu) \nonumber \\
&&+ \frac32({\cal L}_{\bf{\xi}}p+2 \alpha p), \label{lie3}\\
\left[{\cal L}_{\bf{\xi}} \left(T_{\mu\nu} -\frac12 g_{\mu\nu}T\right)\right]h^{\mu\nu} &=& \frac32({\cal L}_{\bf{\xi}}\mu+2\delta \mu) \nonumber \\
&&-\frac32({\cal L}_{\bf{\xi}}p+2\delta p), \\
\left[{\cal L}_{\bf{\xi}} \left(T_{\mu\nu} -\frac12 g_{\mu\nu}T\right)\right] u^\mu h^\nu_{~~\lambda} &=& 0,  \\
\left[{\cal L}_{\bf{\xi}} \left(T_{\mu\nu} -\frac12 g_{\mu\nu}T\right)\right] \left( h^\mu_{~~\lambda} h^\nu_{~~\rho} - \frac{1}{3} h_{\lambda\rho} h^{\mu\nu} \right) &=& 0. \label{lie4}
\end{eqnarray}
Hence, from Eqs.~(\ref{lie1})--(\ref{lie2}) and Eqs.~(\ref{lie3})--(\ref{lie4}), the integrability conditions for the kinematic self-similarity (\ref{intcondition}) are obtained as follows:
\begin{eqnarray} 
(\delta-\alpha)(-8\omega^2 - 2 \dot{u}_\kappa^{~~;\kappa})&=& 8\pi G\left[\frac12({\cal L}_{\bf{\xi}}\mu + 2 \alpha \mu) + \frac32({\cal L}_{\bf{\xi}}p+2 \alpha p)\right],\label{int1}\\
2(\delta-\alpha)(\dot{\theta} + \theta^2 - 4 \omega^2)&=& 8\pi G\left[\frac32({\cal L}_{\bf{\xi}}\mu+2\delta\mu) -\frac32({\cal L}_{\bf{\xi}}p+2\delta p)\right],\label{int2}\\
2 \omega_{\lambda\mu} \dot{u}^\mu+2\omega_{\kappa\lambda}^{~~~;\kappa}-4\omega^2 u_\lambda&=&0,\label{int3}
\end{eqnarray}
and 
\begin{eqnarray}
&&\dot{\sigma}_{\lambda\rho}-u_\rho \sigma_{\lambda\nu} \dot{u}^\nu - u_\lambda \sigma_{\rho\mu} \dot{u}^\mu \nonumber \\
&&~~~~~~~~~~~~~~~+ \theta \sigma_{\lambda\rho} + \sigma_{\lambda\kappa} \omega^\kappa_{~~\rho}+\sigma_{\rho\kappa} \omega^\kappa_{~~\lambda} + 2 \omega_\lambda^{~~\kappa} \omega_{\kappa\rho} + \frac{4}{3} h_{\lambda\rho}\omega^2 =0.\label{int4}
\end{eqnarray} 
These equations must be satisfied in addition to the basic equations. The integrability conditions of the finite kinds (i.e., the first, second and zeroth kinds), assuming `physical' self-similarity for the perfect fluid, i.e.,
\begin{eqnarray}
{\cal{L}}_{\bf\xi} \mu =a\mu,\quad {\cal{L}}_{\bf\xi} p =bp, \label{pss}
\end{eqnarray}
where $a$ and $b$ are constants, were obtained by Coley.~\cite{coley1997} 

If it is assumed that the fluid is irrotational, so that the vorticity is zero (i.e., $\omega_{\mu\nu} =0$), comoving coordinates can be adopted, and we obtain
\begin{eqnarray}
\dot{\mu} + (\mu + p) \theta &=& 0,\label{vortfreebasic1}\\
(\mu + p) \dot{u}_\mu &=& -p_{,\nu} h^\nu_{~~\mu},\label{vortfreebasic2}\\
\dot{\theta} + \frac{1}{3} \theta^2 - \dot{u}^\mu_{~~;\mu}  + 2 \sigma^2 + \frac{8\pi G}{2} (\mu + 3p) &=&0,\label{vortfreebasic3} \\
\left(-\sigma^{\nu\kappa}_{~~~;\kappa} + \frac{2}{3} \theta ^{,\nu}\right) h^\mu_{~~\nu} + \sigma^\mu_{~~\nu}\dot{u}^\nu &=& 0\label{vortfreebasic4}
\end{eqnarray}
from Eqs.~(\ref{b1}), (\ref{b2}), (\ref{b3}), (\ref{b5}) and (\ref{b7}), respectively, while 
\begin{eqnarray} 
- 2(\delta-\alpha) \dot{u}_\kappa^{~~;\kappa}&=& 8\pi G\left[\frac12{\cal L}_{\bf{\xi}}\mu+\alpha\mu + \frac32{\cal L}_{\bf{\xi}}p+3\alpha p\right],\label{vortfreeint1}\\
2(\delta-\alpha)(\dot{\theta} + \theta^2)&=& 8\pi G\left[\frac32 {\cal L}_{\bf{\xi}}\mu+3\delta\mu - \frac32{\cal L}_{\bf{\xi}}p-3\delta p\right],\label{vortfreeint2}\\
0 &=& \dot{\sigma}_{\lambda\rho}-u_\rho \sigma_{\lambda\nu} \dot{u}^\nu - u_\lambda \sigma_{\rho\mu} \dot{u}^\mu+ \theta \sigma_{\lambda\rho} \label{vortfreeint3}
\end{eqnarray} 
from Eqs.~(\ref{int1}), (\ref{int2}) and (\ref{int4}), respectively. The Gauss-Codacci equations must be satisfied in addition. The remaining non-trivial equations serve to define $E_{\mu\nu}$ and $H_{\mu\nu}$. The Einstien tensor on the hypersurface orthogonal to $u^\mu$ are obtained from Eqs.~(\ref{b7}), (\ref{gausseq}), (\ref{codaccieq}) and (\ref{vortfreeint3}) as
\begin{eqnarray}
{}^3R_{\mu\nu} - \frac{1}{3} h_{\mu\nu}{}^3R = E_{\mu\nu} + \sigma_{\mu\kappa} \sigma^\kappa_{~~\nu} - \frac{1}{3} \theta \sigma_{\mu\nu} - \frac{2}{3} \sigma^2 h_{\mu\nu}.\label{3einstein}
\end{eqnarray}

Now we demonstrate the flatness of 3-surfaces orthogonal to the fluid flow for kinematic self-similar dust solutions. When $p=0$, Eq.~(\ref{b2}) gives $\dot{u}_\mu=0$; i.e., the dust fluid is geodesic. From Eq.~(\ref{b3}), we obtain 
\begin{equation} 
\frac{8\pi G}{2}\mu=-\dot{\theta}-\frac{1}{3}\theta^2-2\sigma^2.
\end{equation} 
When $\mu>0$, $\theta$ cannot be identically zero, since $\sigma^2 \ge 0$. Hence the following Lemma is shown.\\
\newtheorem{Lem}{Lemma}
\begin{Lem}
If a spacetime that satisfies the Einstein equations contains an irrotational dust fluid with positive energy density, then $\theta$ cannot be identically zero.
\end{Lem}
\vspace{5mm}
When $p=0$, Eq.~(\ref{vortfreeint1}) gives ${\cal L}_{\bf{\xi}}\mu=-2\alpha \mu$. Hence, the following Lemma is shown.\\
\begin{Lem}
If a spacetime satisfies the Einstein equations, contains an irrotational dust fluid, and admits a kinematic self-similarity vector field of the second, zeroth or infinite kind, then the dust fluid exhibits physical self-similarity with $a=-2\alpha$.
\end{Lem}
\vspace{5mm}
We now show the following theorem.\\
\newtheorem{The}{Theorem}
\begin{The}
Any three surface orthogonal to the fluid flow in a kinemetic self-similar solution to the Einstein equations of the second, zeroth or infinite kind that contains only irrotational dust with positive energy density as a matter field is flat.
\end{The}
\vspace{5mm}
{\it Proof.} Equation (\ref{vortfreeint3}) gives 
\begin{eqnarray}
\dot{\sigma}_{\mu\nu} + \theta \sigma_{\mu\nu} = 0.\label{eq9}
\end{eqnarray}
Multiplying this equation by $\sigma^{\mu\nu}$, we obtain 
\begin{eqnarray}
\sigma[\dot{\sigma} + \theta \sigma] = 0.\label{eq1} 
\end{eqnarray}
Equations (\ref{vortfreeint1}) and (\ref{vortfreebasic3}) with Lemma 2 give
\begin{eqnarray}
(\delta-\alpha)\left(\dot{\theta} + \frac{1}{3} \theta^2 + 2 \sigma^2+8\pi G\frac{\mu}{2}\right) = 0.\label{eq2}
\end{eqnarray}
Differentiating this and using Eqs.~(\ref{eq1}) and (\ref{eq2}), we obtain
\begin{eqnarray}
(\delta-\alpha)\left(\ddot{\theta} + \frac{8}{3} \theta \dot{\theta} + \frac{2}{3} \theta^3+8\pi G\frac{\dot{\mu} + 2 \theta \mu}{2}\right)= 0,\label{eq3}
\end{eqnarray}
which yields, with use of Eq.~(\ref{vortfreebasic1}) and the equation that is obtained by differentiating Eq.~(\ref{vortfreeint2}) with Lemma 2,
\begin{eqnarray}
(\delta-\alpha)\theta\left[\frac23(\dot{\theta} + \theta^2)-8\pi G \mu\right]=0. 
\end{eqnarray}
By Lemma 1, this equation reduces to
\begin{eqnarray}
(\delta-\alpha)\left[\frac23(\dot{\theta} + \theta^2)-8\pi G \mu\right]=0. \label{eq4}
\end{eqnarray}
Hence we obtain 
\begin{eqnarray}
2(\delta-\alpha)\left(- \frac{1}{3} \theta^2 +  \sigma^2+8\pi G\mu\right)=0 \label{eq6}
\end{eqnarray}
from Eqs.~(\ref{eq2}) and (\ref{eq4}). Finally, from Eqs.~(\ref{gausseq}), (\ref{eq9}) and (\ref{eq6}), we consequently find that
\begin{eqnarray}
(\delta-\alpha){}^3R_{\mu\nu} = 0;
\end{eqnarray}
i.e., $3$-spaces orthogonal to $u^\mu$ are flat in the case of kinematic self-similarity of the second, zeroth and infinite kinds, because $\alpha \ne \delta$. \qed \\
\\
If the spacetime is spherically symmetric, the solution can be obtained as the Lema{\^ i}tre-Tolman-Bondi family of solutions. Because the fact that the 3-surfaces orthogonal to the fluid flow are flat corresponds to the fact that these solutions are those for the marginally bound case, we can show the following corollary.\\
\newtheorem{Coro}{Corollary}
\begin{Coro}
A spherically symmetric kinemetic self-similar solution to the Einstein equations of the second, zeroth or infinite kind that contains only dust with positive energy density as its matter field is described by the marginally bound Lema{\^ i}tre-Tolman-Bondi family of solutions.
\end{Coro}

\end{document}